\documentclass[aps,pre,twocolumn,bibnotes,nofootinbib,floatfix]{revtex4-1}
\usepackage{amsmath,amsthm,amsfonts,amssymb,bm}
\usepackage{graphicx,subfigure}
\usepackage{color}
\usepackage{natbib}
\usepackage{hyperref}

\usepackage{mathtools}

\newcommand{\ie}{i.e.,~}
\newcommand{\eg}{e.g.,~}
\newcommand{\secref}{Section~\ref}

\newcommand{\tsr}[1] {\bm{#1}}
\newcommand{\vtr}[1] {\bm{#1}}

\newcommand{\figref}{Fig.~\ref}
\newcommand{\figsref}{Figs.~\ref}
\renewcommand{\eqref}{Eq.~\ref}
\newcommand{\eqsref}{Eqs.~\ref}

\usepackage{epsfig}
\usepackage{epstopdf}

\newcommand{\vecv}[1]{\bm{{#1}}}
\newcommand{\tens}[1]{\bm{{#1}}}

\newcommand{\bi}{\begin{itemize}}
\newcommand{\ei}{\end{itemize}}

\newcommand{\We}{\mathrm{Wi}}
\newcommand{\Weup}{\mathrm{Wi}^{\rm up}}
\newcommand{\Wezero}{\mathrm{Wi}_0}
\newcommand{\Weone}{\mathrm{Wi}_1}
\newcommand{\Wetwo}{\mathrm{Wi}_2}
\newcommand{\Wemax}{\mathrm{Wi}_{\rm max}}
\newcommand{\Wec}{\We_\textrm{c}}

\newcommand{\Rcross}{R_{\rm c}}

\newcommand{\gdot}{\dot{\gamma}}

\newcommand{\be}{\begin{equation}}
\newcommand{\ee}{\end{equation}}
\newcommand{\beqna}{\begin{eqnarray}}
\newcommand{\eeqna}{\end{eqnarray}}

\newcommand{\edot}{\dot{\epsilon}}

\newcommand{\gradv}{\nabla v}

\newcommand{\etal}{\emph{et al. }}

\newcommand{\lambdad}{\lambda_{\tens{D}}}
\newcommand{\lambdaconf}{\lambda_{\tens{W}}}
\newcommand{\lambdao}{\lambda_{\tens{\Omega}}}

\newcommand{\total}{\tens{T}}
\newcommand{\visc}{\tens{\Sigma}}
\newcommand{\conf}{\tens{W}} %

\begin{document}

\title{Thickening of viscoelastic flow in a model porous medium}

\author{E. J. Hemingway$^1$, A. Clarke$^2$, J. R. A. Pearson$^2$ and S. M. Fielding$^1$\\
$^1$Department of Physics, Durham University, Science Laboratories, South Road, Durham, DH1 3LE, UK\\
$^2$Schlumberger Gould Research, Madingley Road,
  Cambridge, UK, CB3 0EL, UK\\}

\date{\today}
\begin{abstract}
  We study numerically two-dimensional creeping viscoelastic flow past
  a biperiodic square array of cylinders within the Oldroyd B, FENE-CR
  and FENE-P constitutive models of dilute polymer solutions. Our
  results capture the initial mild decrease then dramatic upturn
  (`thickening') seen experimentally in the drag coefficient as a
  function of increasing Weissenberg number. By systematically varying
  the porosity of the flow geometry, we demonstrate two qualitatively
  different mechanisms underpinning this thickening effect: one that
  operates in the highly porous case of widely spaced obstacles, and
  another for more densely packed obstacles, with a crossover between
  these two mechanisms at intermediate porosities.  We also briefly
  consider 2D creeping viscoelastic flow past a linear array of
  cylinders confined to a channel, where we find that the flow is steady for
  all Weissenberg numbers explored.

\end{abstract}

\maketitle

\section{Introduction}

Flows of polymeric melts and solutions exhibit a rich phenomenology that has been widely studied \cite{Larson2014}.  Many such materials exhibit viscoelastic properties which dramatically affect their behaviour in processes used in their industrial application.  Examples range from melt extrusion through coating process to flow in porous materials.  The latter example has particular relevance in the oil and gas industry where viscoelastic solutions are routinely employed for enhanced oil recovery, matrix stimulation and fracturing.  For enhanced oil recovery, aqueous polymer solutions have long been used to control the Saffman-Taylor instability \cite{Homsy1987,Bensimon1986} that results from the displacement of viscous oil by lower viscosity brine within a porous reservoir rock.  Since the earliest experiments it has been known that for certain polymeric solutions, even for single-phase flow, an anomalously high pressure gradient is observed when compared with the flow of an appropriate equivalent Newtonian solution.  Whereas this phenomenon has been long known, no clear understanding of the root cause of the high observed drag has been forthcoming.  In particular there has been no clear elucidation of the relative contributions of shear, extensional and elastic stresses.  The excess pressure gradient directly impacts the industrial process by limiting the rate of injection of polymeric solution, where an excessive pressure gradient will lead to fracturing of the rock in the vicinity of the injector and to degradation of the polymer.

Experimentally, several authors have studied the flow of a viscoelastic fluid past a biperiodic array of obstacles arranged in two spatial dimensions~\cite{Chmielewski1992,Skartsis1992,Khomami1997,Howe2015,James2012,James2016}.  (The upper panel of Fig.~\ref{fig:geometries} shows a sketch of the simplified cartoon of such a geometry that we shall study numerically in this work.)  The dominant physical effect consistently reported in these experiments is a dramatic upturn in the adimensional pressure drop, known as the drag coefficient, relative to that for a Newtonian fluid of matched viscosity, for Weissenberg numbers $\We$ exceeding a critical value $\Wec$. (The Weissenberg number is the product of a characteristic shear-rate in the flow and polymer relaxation time $\tau$. Later in the text we define three Weissenberg numbers suited to the problem at hand.) This upturn is often also accompanied by the development of time-dependent flow fields, with crossing streaklines and structure in the third spatial dimension~\cite{Chmielewski1992,Khomami1997}, into the page in the simplified sketch of Fig.~\ref{fig:geometries}. For low Weissenberg numbers a subtler effect is sometimes also seen, in which the pressure drop initially decreases slightly relative to the Newtonian case~\cite{Khomami1997} before the upturn just described for $\We > \Wec$.

Some of these studies \cite{Clarke2015,Clarke2015a} correlate the observed upturn in the drag coefficient (described therein via an apparent viscosity) found for polymer solution flow in outcrop rock samples with flows of the same solutions observed in 2D microfluidic networks.  These networks comprise ``pores'' and randomly sized ``throats'' on a grid rotated $45^\circ$ to the average flow direction. The experiments were subsequently extended \cite{Howe2015} to show how the onset of additional drag depended on solution parameters.  In particular, the onset of thickening was found to be well characterized by a Weissenberg number derived using a characteristic apparent shear rate proportional to flow rate together with a molecular relaxation time (see Fig.~11 in that work).  In these studies, the transition to a time-dependent state resembling turbulence is marked by the appearance of crossing streaklines, a 3D effect which isn't possible in the current 2D study.  Nuclear magnetic resonance studies \cite{Mitchell2016} demonstrated time-dependent flows within a full 3D pore network (a rock) via an effective diffusion constant measurement. The analysis developed in the following paper can also be applied to these geometries at 45 degree orientations. Our results (not discussed here) suggest that the cylinder size has less effect on the flow character $q$ (defined in \secref{sec:character}) than for $0^\circ$ orientations. However we defer a full study in these rotated geometries to future work.

From a theoretical viewpoint, early attempts to understand flow in porous media adopted a coarse-grained approach, discarding microscopic details in favour of macroscopic properties. For Newtonian flow, Darcy \cite{darcy1856} proposed a relation between the pressure drop per unit length of material $\Delta P / L$ and the mean velocity scale $V$
\be
\frac{\Delta P}{L} = \frac{\eta V}{K},
\label{eq:porous_intro_darcy}
\ee
where $\eta$ is the fluid's viscosity. The permeability $K$ is a
constant that should depend only on the properties of the medium, but
is unknown {\it a priori}. A relation between the permeability and the
medium's porosity $\epsilon$ (the ratio of free volume to total
volume) was later proposed in the Blake-Kozeny-Carman equation
\cite{Carman1937}, which proved successful in describing a range of
simple flows. However the validity of this macroscopic approach
remains largely limited to Newtonian flows.

In theoretically understanding viscoelastic creeping flows in porous
media, much of the progress has been made
computationally~\cite{Talwar1992,Talwar1995,Souvaliotis1996,LiuThesis1997,Alcocer1999,Alcocer2002,Gillissen2013}.
Early simulations of two-dimensional (2D) viscoelastic flow past a
bi-periodic square~\cite{Talwar1992,Talwar1995}
or rectangular~\cite{Souvaliotis1996}
array of cylinders observed a slightly reduced pressure drop at low
Weissenberg numbers compared to that of a Newtonian fluid of matched
viscosity, as seen experimentally. However the dramatic upturn in the
drag coefficient at high $\We>\Wec$, which is the dominant physical
effect seen experimentally, was not captured in these early numerical
works, presumably due to the restricted computational processing power
available at the time.  It was however later captured in simulations
of the Oldroyd B and FENE-CR models, also in 2D biperiodic
arrays~\cite{LiuThesis1997}.

Alcocer et al.~\cite{Alcocer1999,Alcocer2002} investigated 2D flow of
the FENE-CR model past a biperiodic array of cylinders, with a
particular focus on the dependence of the effective permeability on
the cell aspect ratio, for a fixed area fraction of cylinders. They
demonstrated a non-monotonic dependence of permeability on aspect
ratio.  For a fixed aspect ratio and area fraction, they reported in
an initial increase in permeability with increasing $\We$, equivalent
to the initial decrease in drag in other studies.

Gillissen~\cite{Gillissen2013} simulated 2D flow of the FENE-P model past a biperiodic hexagonal array of cylinders.  This study convincingly captured both the initial downturn and then significant upturn in the drag coefficient seen experimentally as a function of Weissenberg number. Analysing the flow field as a function of space in terms of regions of pure shear (which is the same as extension), simple shear and pure rotation, they demonstrated a predominance of shear regions at low $\We$, with a progressive increase in elongational regions with increasing $\We$. At high $\We$ the polymer conformation tensor was found to be fully extended, showing the importance of finite chain extensibility in this regime.

De \etal studied 3D flow of a FENE-P fluid past an array of cylinders (both with and without walls) \cite{De2016}. They found an elastic instability whereby recirculating regions in the cylinder wake break symmetry and form a 3D structure, which occurs at a Deborah number $De = \tau V / R$ consistent with Ref.~\cite{Smith2003}. All their simulation runs attained a time-independent steady-state.

Besides the biperiodic geometries just discussed, significant efforts have also been devoted to understanding viscoelastic flow past a single cylinder or linear array of cylinders confined to a channel: experimentally~\cite{McKinley1993,LiuThesis1997,Moss2010a,Ribeiro2014,Zhao2016}, by linear stability analysis~\cite{Smith2003,Sahin2008}, and by direct numerical simulation~\cite{Ribeiro2014,Liu1998a,Hulsen2005,Oliveira2005,Claus2013,LiuThesis1997,Vazquez-Quesada2012,Grilli2013}.  The lower panel of Fig.~\ref{fig:geometries} shows a sketch of the simplified cartoon of such a geometry that we shall study numerically in this work.

By studying the flow of Boger fluids past a single cylinder in a channel, McKinley \etal \cite{McKinley1993} observed a transition from steady 2D to steady 3D flow in the downstream wake. At higher flow rates, they found another instability where time-dependent velocity oscillations form in the wake region. Liu \cite{LiuThesis1997} considered flow past a single cylinder, and widely and closely spaced linear arrays of cylinders. For the single cylinder, they observed a mild downturn in the drag at moderate $\We$, followed by an upturn at larger $\We$ which was accompanied by a transition from steady 2D to steady 3D flow. In the linear arrays, the transition was from steady 2D to time-dependent 3D structure in both cases. Moss and Rothstein \cite{Moss2010a} studied flow of wormlike-micelles (WLMs) past a single cylinder for several ratios of cylinder diameter to channel width. They observe a significant decrease in the normalised pressure drop as a function of $\We$ which was attributed to the shear-thinning properties of the fluids. Of the two fluids tested, only one exhibited an instability, which was attributed to breakdown of the WLMs in the extensional flow in the cylinder wake. Using flow-induced birefringence measurements, the authors showed that shear flows at the channel walls were not necessary to produce the wake instability. Recent experiments on WLMs \cite{Zhao2016} found that upstream vortices formed at much larger $\We \sim 10^3$, and unsteady flow downstream at even larger $\We \sim 10^4$.

Simulations of 2D creeping viscoelastic flow past a linear array of cylinders confined in a channel~\cite{Liu1998a,Hulsen2005} captured an initial mild decrease then subsequent upturn in the drag coefficient as a function of increasing Weissenberg number. All the states observed were however steady as a function of time.  In 2D simulations of flow past a single cylinder confined in a channel, temporal oscillations in the size of a recirculating region that forms downstream of the cylinder are seen~\cite{Oliveira2005,Claus2013}, with an associated slight increase in the drag coefficient compared with the time-independent state.

Attempts to understand the onset at high $\We$ of the 3D time-dependent states seen experimentally for creeping viscoelastic flow past an array of cylinders in a channel have been made by performing a linear stability analysis for the dynamics of small amplitude 3D perturbations to an initially 2D flow state.  By such an analysis, Smith et al. \cite{Smith2003} reproduced some of the experimentally observed characteristics of the instability, in particular the wavevector of the most unstable mode and the critical $\We$ at which the instability first arises.  Sahin and Wilson~\cite{Sahin2008} demonstrated that the wavelength of the instability scales with cylinder spacing for closely spaced cylinders, and with the size of the wake behind the cylinder for wider cylinder spacing.  Both studies warn that this 3D instability potentially restricts the range of $\We$ over which purely 2D simulations might remain valid. Also we stress that linear stability does not preclude the existence of nonlinear instabilities, \eg as discussed by Pan \etal \cite{Pan2013}.

V\'{a}zquez-Quesada and Ellero performed 2D smoothed particle hydrodynamics (SPH) simulations of viscoelastic flow past an array of cylinders in a channel \cite{Vazquez-Quesada2012,Grilli2013}, capturing the initial downturn and subsequent upturn in the drag coefficient as a function of Weissenberg number, as seen experimentally.  At higher $\We$ the results of this study departed from other 2D numerical works \cite{LiuThesis1997} in reporting a transition to a time-dependent state, which the authors interpreted as viscoelastic turbulence.  We do not find this viscoelastic turbulence, and discuss carefully the differences between our study and Refs.~\cite{Vazquez-Quesada2012,Grilli2013} that might potentially explain this apparent discrepancy between the two studies.

Ribeiro et al.~\cite{Ribeiro2014} studied fully 3D viscoelastic flow past a cylinder confined to a channel, both experimentally and numerically, for both a shear thinning and a Boger fluid. For the shear-thinning fluid they reported an elastic instability setting in upstream of the cylinder at critical value of $\We$ that depends on the cylinder height in the vorticity direction. For Weissenberg numbers just beyond onset of the instability, the system's state is asymmetric and time-independent. A subsequent transition to a time-dependent state was reported at larger $\We$ still.

Among the simulation studies of viscoelastic flow just surveyed~\cite{Talwar1992,Talwar1995,Souvaliotis1996,LiuThesis1997,Alcocer1999,Alcocer2002,Gillissen2013}, each considered one (or in some cases two) fixed ratio(s) of obstacle size to obstacle spacing, \ie a fixed medium porosity. A key contribution of this work is systematically to vary the medium porosity over a broad range, from the limit of widely spaced obstacles to ones that nearly touch. In doing so, we shall demonstrate two qualitatively different mechanisms underpinning the dramatic upturn of the drag coefficient with Weissenberg number seen experimentally: one at low obstacle area fraction and another at high area fraction, with a crossover between these mechanisms at intermediate area fraction.

In neither case, however, do we find the thickening to be associated with the onset of a time-dependent flow, as often reported experimentally. The same is true in the 2D biperiodic simulations in the earlier literature~\cite{Talwar1992,Talwar1995,Souvaliotis1996,LiuThesis1997,Alcocer1999,Alcocer2002,Gillissen2013}.  One possible explanation for this discrepancy is that the time-dependent state seen experimentally is 3D in nature, and cannot be captured in purely 2D simulations.

The paper is structured as follows.  In Secs.~\ref{sec:geometries} and~\ref{sec:models} we introduce the flow geometries and constitutive models to be studied.  The governing parameters and dimensionless groups are summarised in Sec.~\ref{sec:parameters}. In Sec.~\ref{sec:numerics} we outline our numerical methods, and provide benchmarks to validate them against known results in the Newtonian limit. We then present our results for viscoelastic flow: in Sec.~\ref{sec:biperiodic} for a biperiodic array of cylinders and in Sec.~\ref{sec:channel} for a linear array of cylinders in a channel.

\section{Flow geometries}
\label{sec:geometries}

\begin{figure}[t]
  \includegraphics[width=\columnwidth]{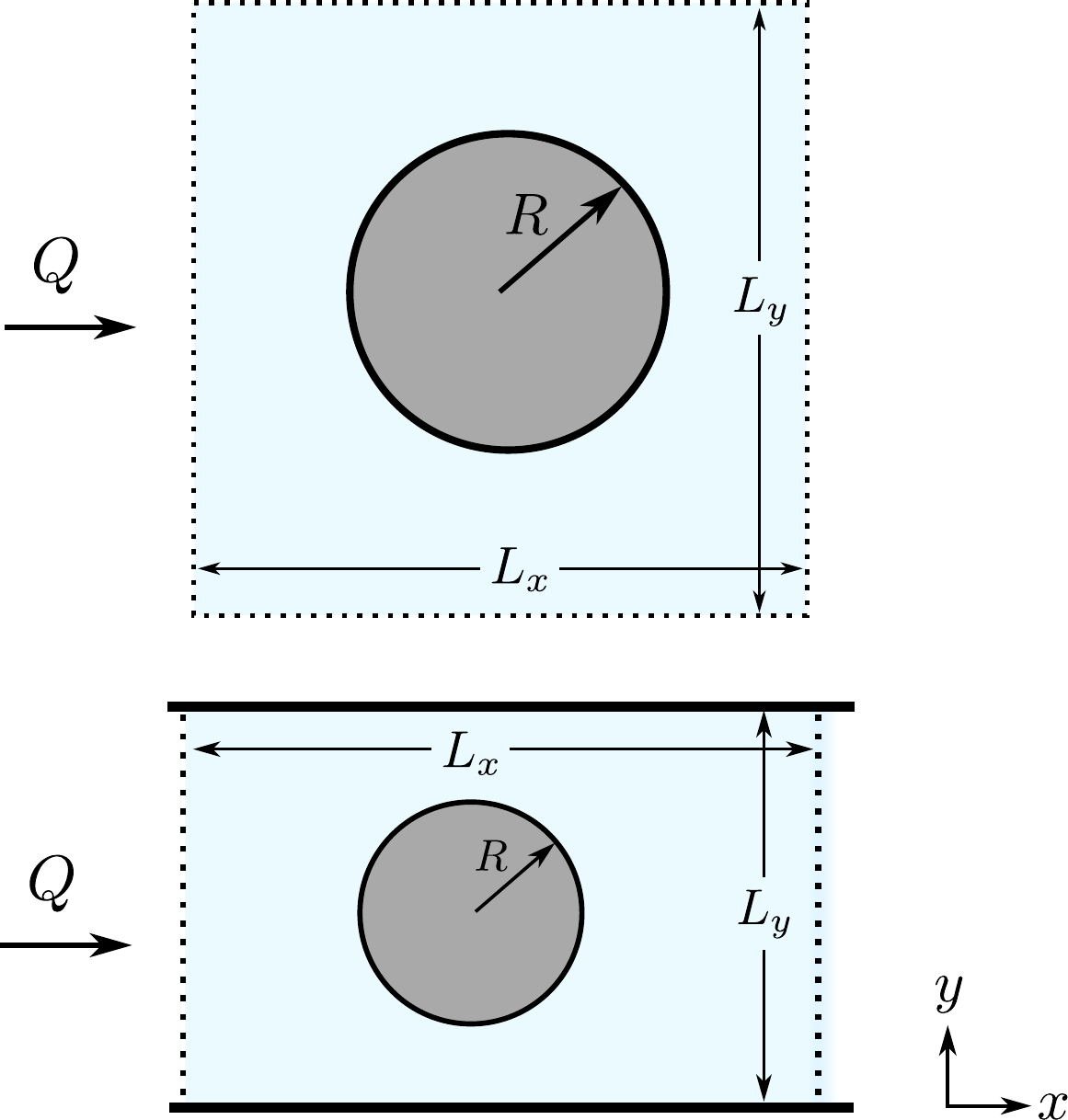}
  \caption{Flow geometries to be studied. {\bf Upper:} Biperiodic array of cylinders. {\bf Lower:} Linear array of cylinders in a channel.  Solid lines represent closed walls, and dashed lines represent periodic boundaries. The flow is assumed translationally invariant into the page.}
\label{fig:geometries}
\end{figure}

We shall study two different flow geometries. The first comprises a biperiodic array of cylinders, sketched in the upper panel of Fig.~\ref{fig:geometries}.  The second comprises a linear array of cylinders in a channel bounded by solid walls, sketched in the lower panel of the same figure.  In each case the flow cell has length $L_x$ horizontally and height $L_y$ vertically, and the cylindrical obstacle has radius $R$. At the boundaries of the cells represented by dashed lines all flow variables are ascribed periodic boundary conditions. At the cell boundaries represented by solid lines, and at the cylinder surface, conditions of no-slip and no-permeation apply. In each case we assume the flow to be translationally invariant into the page in Fig.~\ref{fig:geometries}, and simulate the flow in the two dimensions of the page only.

The flow, which we assume to be from left to right, can be imposed
in two different ways. In the first, a given throughput $Q$ per unit
time is prescribed (such that the characteristic velocity scale is
then $V=Q/L_y$), with the pressure drop $\Delta P$ measured in
response.  Alternatively, we can impose the pressure drop $\Delta P$
and measure the resulting throughput $Q$.

In either case, a key experimental observable is the drag coefficient
\be
\label{eqn:drag}
C_D=\frac{\Delta P L_y}{\eta V},
\ee
which measures the pressure drop normalised by the characteristic
velocity scale and the solvent viscosity. In the limit of Newtonian
flow this quantity depends only on the flow geometry. In non-Newtonian
flow it also depends via the Weissenberg number on the nonlinear
constitutive behaviour of the fluid in question. In reporting our
numerical results below we shall typically show the drag coefficient
at any given Weissenberg number, $C_D(\We)$, normalised by the
corresponding value in the Newtonian limit of zero Weissenberg number,
$C_D(\We\to 0)$, defining
\be
\label{eqn:dragnormed}
\chi=\frac{C_D(\We)}{C_D(\We\to 0)}.
\ee

\section{Constitutive models}
\label{sec:models}

We write the total stress $\total(\tens{r},t)$ in a fluid element at
position $\tens{r}$ and time $t$ as the sum of a viscoelastic
contribution $\visc(\tens{r},t)$ from the polymer chains, a Newtonian
solvent contribution of viscosity $\eta$, and an isotropic
contribution with a pressure $p(\tens{r},t)$:
\be
\total = \visc + 2 \eta \tens{D} - p\tens{I}.
\label{eqn:total_stress_tensor}
\ee
The symmetric strain rate tensor $\tens{D} = \frac{1}{2}(\tens{\gradv} +
\tens{\gradv}^T)$ where $\tens{\gradv}|_{\alpha\beta} =
\partial_{\alpha}v_{\beta}$ and $\tens{v}(\tens{r},t)$ is the fluid
velocity field.

Throughout we consider the creeping flow limit of zero Reynolds
number. Here the condition of force balance requires the stress field
$\total(\tens{r},t)$ to be divergence free:
\be
\vecv{\nabla}\cdot\,\total = 0,
\label{eqn:force_balance}
\ee
such that
\be
\label{eqn:Stokes}
 \nabla.\tens{\Sigma}+\eta\nabla^2\vecv{v}-\nabla p = 0.
\ee
The pressure field $p(\tens{r},t)$ is determined by enforcing flow
incompressibility:
\be
\label{eqn:incomp}
\vecv{\nabla}\cdot\vecv{v} = 0.
\ee

The dynamics of the polymeric stress $\visc$ is specified by a viscoelastic constitutive model. In this work we consider three different phenomenological constitutive equations: the Oldroyd B, FENE-CR and FENE-P models~\cite{Larson1988}. These each describe a dilute polymer solution by representing each polymer chain as a simplified dumbbell comprising two beads connected by a spring.  The conformation tensor $\conf=\langle \vecv{R}\vecv{R}\rangle$ is defined as the ensemble average $\langle \rangle$ of the outer dyad of the dumbbell end-to-end vector $\vecv{R}$, which is taken to have unit length in equilibrium.  The conformation of the polymer chains determines the viscoelastic stress according to
\be
\label{eqn:stress}
\visc = G\left[f(\conf) \conf - g(\conf) \mathbf{I} \right],
\ee
with a constant modulus $G$.  In addition to the spring force, each
bead also experiences viscous drag against the
solvent~\cite{Larson1988} and stochastic thermal fluctuations.  The
conformation tensor is then taken to obey
\be
\overset{\nabla}{\conf} = -\frac{1}{\tau} \left[ f(\conf)\conf - g(\conf) \mathbf{I} \right] + \frac{\ell^2}{\tau} \nabla^2 \conf,
\label{eqn:Maxwell}
\ee
with a characteristic relaxation time $\tau$, where
\begin{equation}
  \overset{\nabla}{\conf} \equiv  \left(\partial_t + \vtr{v}\cdot \nabla \right) \conf - \tsr{\gradv}^T\cdot\conf- \conf\cdot\tsr{\gradv}
\end{equation}
is the upper convected derivative~\cite{Larson1988}.  We have included a modification to the original equations by introducing a diffusive term, where $\ell$ is a small lengthscale below which gradients in $\conf$ are attenuated. Similar modifications have been made to the Johnson-Segalman model in the context of shear banding \cite{Johnson1977,Radulescu1999}. Specifically in the context of porous media, Gillissen included diffusive terms in his study of a FENE fluid past a biperiodic array of cylinders \cite{Gillissen2013}. Thomases \etal also recently showed that a small diffusive contribution can support a finite polymer stress in a qualitatively similar fashion to FENE models \cite{Thomases2007,Thomases2011}.  Without diffusion included, we find that the severe space- and time-step requirements limit the range of Weissenberg numbers that can feasibly be explored.

Gradient terms in $\conf$ require a boundary condition at the walls and at the cylinder surface. For simplicity we choose zero gradient at the walls $\partial_y \conf = 0$ (when present). While the zero gradient condition can also explicitly be imposed on the cylinder surface\footnote{The method is analogous to the procedure used to enforce no-slip boundary conditions on the cylinder surface, see the discussion preceding Eq.~\ref{eqn:porous_tether}. Essentially, we add a term to restore the gradient normal to the cylinder to zero.}, we find that our numerical method (which solves for $\tsr{W}$ everywhere in the computational domain, both inside and outside of the cylinder) in fact naturally produces an emergent zero gradient at a typical radius $R + \ell$. We therefore find that it is sufficient to ensure that $\ell$ is small compared to any other physical lengthscale in the problem:  we fix $\ell = 0.01$ in all that follows and we have checked that our results are qualitatively (and in almost all cases quantitatively\footnote{The only exception is in the closely spaced cylinders case shown in \figref{fig:drag_Qc_Qd} (right) where the drag upturn is slightly less steep for $\ell = 0.005, 0.0025$. The point of upturn in the drag is unchanged in all cases, and all states remain time-independent.  }) unchanged by further decreasing $\ell$.

\begin{figure}[tbp]
  \includegraphics[width=\columnwidth]{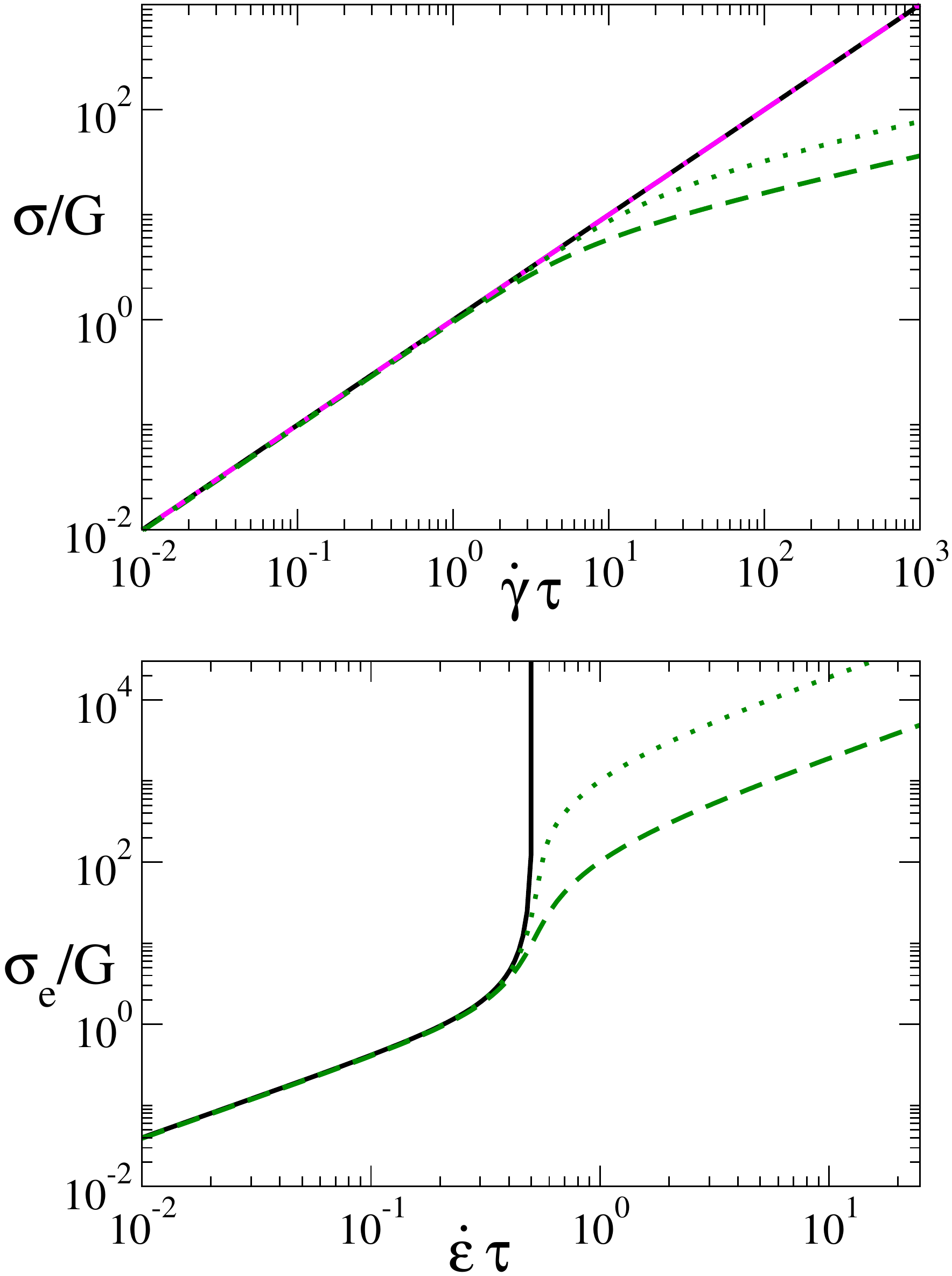}
  \caption{{\bf Upper:} Stationary constitutive curves for homogeneous simple shear flow in the Oldroyd B model (solid black curve), the FENE-CR for $\delta=0.001$ (magenta dotted curve) and $0.01$ (magenta dashed curve) and the FENE-P model for $\delta=0.001$ (green dotted curve) and $\delta=0.01$ (green dashed curve).  The FENE-CR curves are indistinguishable from the Oldroyd B curve.  \newline {\bf Lower:} Counterpart stationary constitutive curves for homogeneous planar extensional flow, for the same parameter values as in the upper panel, and with the same line key. In this case, for matched $\delta$, the FENE-CR and FENE-P curves are indistinguishable from each other (and the two magenta curves are accordingly hidden by the green ones).}
\label{fig:constitutive}
\end{figure}

The functions $f(\conf)$ and $g(\conf)$ in Eqns.~\ref{eqn:stress} and~\ref{eqn:Maxwell} are different in the three different models. The Oldroyd B model has $f(\conf)=g(\conf)=1$, which corresponds to assuming that the spring of each dumbbell is Hookean.  For a sustained imposed extensional strain rate $\edot> 1/2\tau$ this model displays an extensional catastrophe in which the dumbbells stretch out indefinitely and the extensional stress grows indefinitely.

The phenomenological FENE-CR and FENE-P models regularise this catastrophe by insisting that the extensional stress of the polymer chains (dumbbells) must remain finite at all deformation rates.  They do so by replacing the Hookean spring law by a non-linear law with finite-extensibility \cite{Chilcott1988}.  The FENE-CR model has
\be
\label{eqn:feneCR}
f(\conf)=g(\conf)=\alpha(\conf)\equiv\frac{1}{1-\frac{Tr(\conf)}{\Lambda^2}},
\ee
while the FENE-P model has
\be
\label{eqn:feneP}
f(\conf)=\alpha(\conf)\;\;\rm{and}\;\; g(\conf)=1.
\ee

The parameter $\Lambda$ in Eqn.~\ref{eqn:feneCR} characterises the maximum
extent to which any dumbbell can be stretched. We choose to express
this in terms of $\delta= \Lambda^{-2}$, noting that the limit $\delta\to 0$
corresponds to Oldroyd B dynamics with infinite extensibility.

Under conditions of ideal viscometric simple shear flow, the imposed
velocity gradient tensor in Cartesian ($x-y$) coordinates is
\be
\tens{\gradv}|_{\rm shear}=\gdot \left(\begin{matrix}
    0       & 0 \\
    1       & 0
\end{matrix}\right).
\label{eqn:shear}
\ee
In steady state under this applied flow, the shear stress $\Sigma_{xy}$ (which we also denote $\sigma$) as a function of $\gdot$ defines the fluid's shear constitutive curve, as shown in Fig.~\ref{fig:constitutive} for the three models considered here. The shear viscosity of the Oldroyd B and FENE-CR models is constant as a function of $\gdot$.  The FENE-P model shear thins to an extent determined by the value of $\delta$.

In an ideal viscometric planar extensional flow, the imposed velocity
gradient tensor
\be
\tens{\gradv}|_{\rm ext}=\edot \left(\begin{matrix}
    1       & 0 \\
    0       & -1
\end{matrix}\right).
\label{eqn:ext}
\ee
In steady state, the extensional stress $\Sigma_{xx}-\Sigma_{yy}$
(which we also denote $\sigma_{\rm E}$) as a function of $\edot$
defines the fluid's extensional constitutive curve, as shown in
Fig.~\ref{fig:constitutive} for the three models. As can be seen, the
extensional constitutive curve of the Oldroyd B model is undefined for
flow rates $\edot>1/2\tau$, consistent with our discussion above of
the chain stretch catastrophe.  In contrast, the FENE models have
well-defined constitutive curves at all strain rates $\edot$. Indeed
for matched $\delta$ the two FENE models have the same extensional
constitutive curves, with a finite limiting extensional viscosity $2
G\tau/\delta$ as $\edot\to\infty$.

\section{Parameters and dimensionless groups}
\label{sec:parameters}

The eight parameters characterising the fluid, geometry, and imposed
flow just described are: the solvent viscosity $\eta$, the polymer
modulus $G$, the polymer viscoelastic relaxation timescale $\tau$, the
polymer finite extensibility parameter $\delta$, the cell length
$L_x$, the cell height $L_y$, the cylinder radius $R$ and the flow's
throughput rate $Q$.

We are free to choose units of mass, length and time, leaving five
dimensionless groups as follows: the ratio of cylinder radius to gap
height $\tilde{R}=R/L_y$, the ratio of cell length to cell height
$\tilde{L}_x=L_x/L_y$, the ratio of solvent to total (zero shear)
viscosity $\beta=\eta/\left(\eta + G\tau\right)$, the finite extensibility parameter
$\delta$, and a Weissenberg number $\We$ characterising the strength
of the velocity gradients compared to the inverse of the fluid's
stress relaxation time $\tau$. (We return below to define $\We$
precisely.)  We drop tildes hereafter with the understanding that (for
example) values for $R$ and $L_x$ are always quoted in units of $L_y$.
To allow benchmarking of our results against the earlier literature we
fix the viscosity ratio $\beta=0.59$ throughout.  Remaining to be
explored numerically are then the dimensionless cylinder radius $R$,
the dimensionless cell length $L_x$, the finite extensibility
parameter $\delta$ and the Weissenberg number $\We$.

The Weissenberg number is a dimensionless quantity characterising the
scale of velocity gradients in the flow in units of the inverse
relaxation time. On purely dimensional grounds, one simple possible
definition is $V\tau/L_y$. However we have found it more useful to
define two different Weissenberg numbers based on a characterisation
of the velocity gradients that actually develop inside the flow
geometry.  We shall return to discuss these in
Sec.~\ref{sec:biperiodic} below.

In our studies of the biperiodic geometry we consider a square cell
with $L_x= L_y \equiv L = 1.0$. For the channel geometry we take the cylinder spacing
$L_x$ as an additional variable to be explored numerically.

\section{Numerical methods}
\label{sec:numerics}

We numerically solve the model equations~\ref{eqn:Stokes},~\ref{eqn:incomp},~\ref{eqn:stress} and~\ref{eqn:Maxwell} using a timestepping approach. Each main timestep comprises two separate substeps, as follows.  In the first substep the viscoelastic stress $\tens{\Sigma}(\vecv{r},t)$ is updated according to the viscoelastic constitutive equations~\ref{eqn:stress} and~\ref{eqn:Maxwell} with a fixed velocity field $\vecv{v}(\vecv{r},t)$.  This update is performed in the whole plane of Fig.~\ref{fig:geometries}, even inside the cylinder, where the velocity is zero to within the accuracy of our numerical approach.  For this substep we adopt a method that we have used in our own previous work~\cite{Fielding2006}, to which we refer the reader for details.

With that newly updated viscoelastic stress field, the velocity field
is then updated in the second substep to satisfy
Eqns.~\ref{eqn:Stokes} and~\ref{eqn:incomp}. This substep needs more
careful discussion, because the presence of the cylindrical posts
renders the procedure more complicated than in
Ref.~\cite{Fielding2006} (which considered a rectangular flow cell
with no obstacles). In particular, the boundary conditions of no-slip
and no-permeation must be satisfied for the fluid velocity round the
edge of each obstacle. To tackle this we use an immersed boundary
method (IBM)\cite{Peskin1972,Peskin2002,Mittal2005}, which couples a
solution of Eqns.~\ref{eqn:Stokes} and~\ref{eqn:incomp} on a regular
Eulerian grid with an off-grid Lagrangian description of the cylinder
surface.  (We use the term Lagrangian to be consistent with the IBM
literature, although in our particular problem the cylinder doesn't
advect with the flow.)

The cylinder surface is characterised by a one-dimensional curvilinear
Lagrangian coordinate $\xi$ around it. (Although we use the word
cylinder, in our 2D study it is of course represented by a circular
cross section, with a 1D edge.) The location of a point at the
Lagrangian coordinate $\xi$ is given in the Eulerian frame as
$\vecv{X}(\xi, t)$. A Lagrangian force density $\vecv{F}(\xi, t)$ is
then incorporated, calculated by prescribing the desired location of
the cylinder $\vecv{X}_0(\xi)$, and imposing a Hookean restoring force
when the cylinder deviates from this:
\be
  \vecv{F}(\xi, t) = -\kappa \left(\vecv{X}(\xi, t) - \vecv{X}_0(\xi)\right),
  \label{eqn:porous_tether}
\ee
where $\kappa$ is a large spring constant. Given that the desired
location $\vecv{X}_0$ is independent of time, differentiating
Eqn~\ref{eqn:porous_tether} with respect to time gives
\be
  \frac{\partial \vecv{F}(\xi, t)}{\partial t} = -\kappa \vecv{U}(\xi, t),
  \label{eqn:porous_tether_mod}
\ee
in which the Lagrangian velocity $\vecv{U}$ at the cylinder surface is
calculated from the velocity $\vecv{v}(\vecv{r},t)$ in the Eulerian
frame as
\be
  \vecv{U}(\xi, t)  = \int_{\Omega} \vecv{v}(\vecv{r}, t) \delta(\vecv{r} - \vecv{X}(\xi, t)) d\vecv{x}.
  \label{eqn:porous_eul_to_lagr}
\ee

Eqn.~\ref{eqn:porous_tether_mod} is evolved at each timestep using an
explicit Euler algorithm. The Lagrangian force density $\vecv{F}(\xi,
t)$ then gives a force contribution in the Eulerian frame of
\be
  \vecv{f}(\vecv{r}, t) = \int_{\Omega} \vecv{F}(\xi, t) \delta(\vecv{r} - \vecv{X}(\xi, t)) d\xi.
  \label{eqn:porous_lagr_to_eul}
\ee
This is incorporated as an additional source term to the left hand side of the generalised Stokes' equation~\ref{eqn:Stokes}, which is then solved on a rectangular Eulerian grid in the full plane of Fig.~\ref{fig:geometries} using the methods of Ref.~\cite{Fielding2006}. Without walls, this can be done either imposing the overall throughput (and measuring the pressure drop) or imposing the pressure
drop (and measuring the throughput), and we have checked that these are
equivalent in all our biperiodic results presented below. With walls included, we always impose the latter quantity.

\begin{figure}[tbp]
  \centering
  \includegraphics[width=\columnwidth]{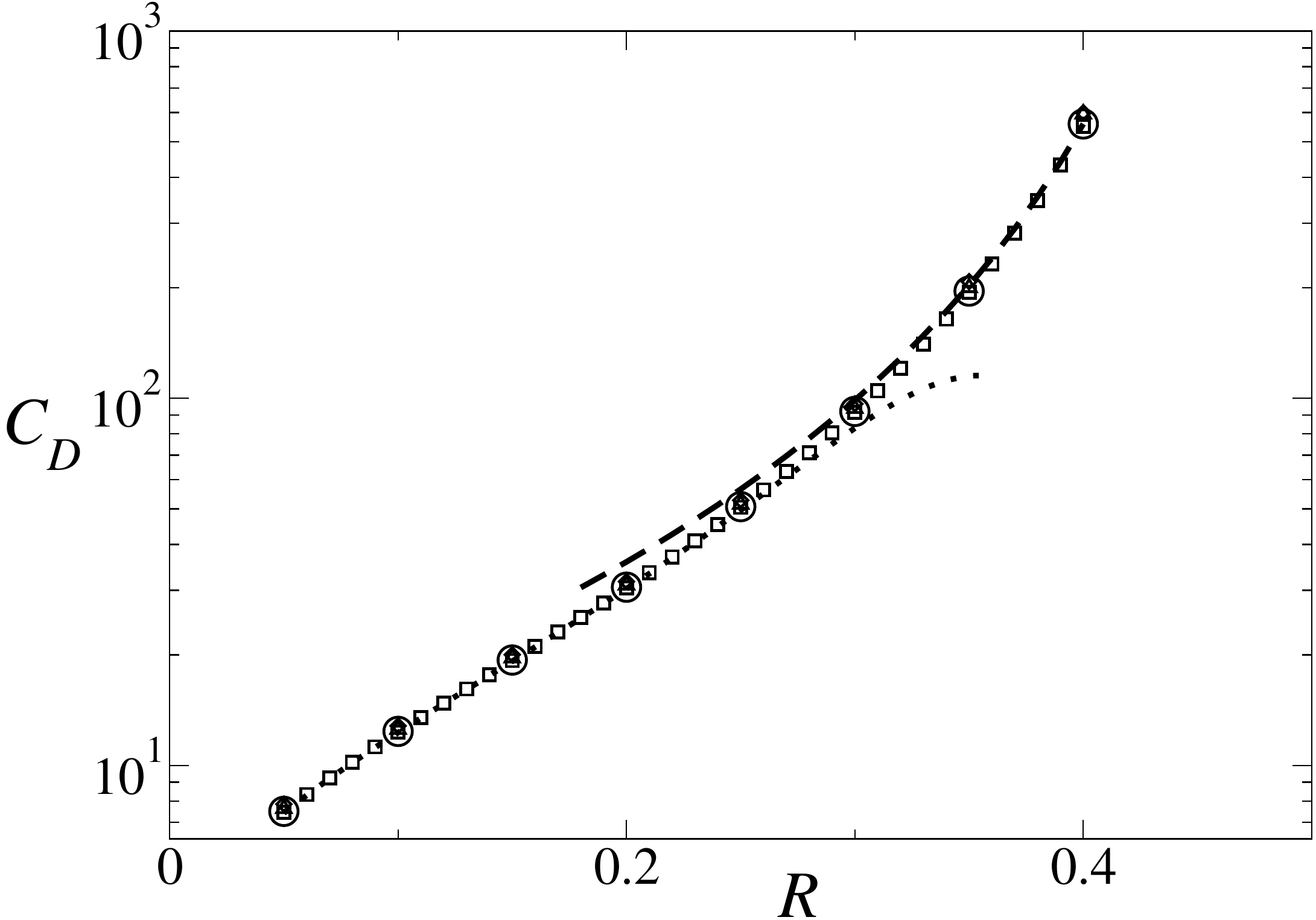}

  \includegraphics[width=0.95\columnwidth]{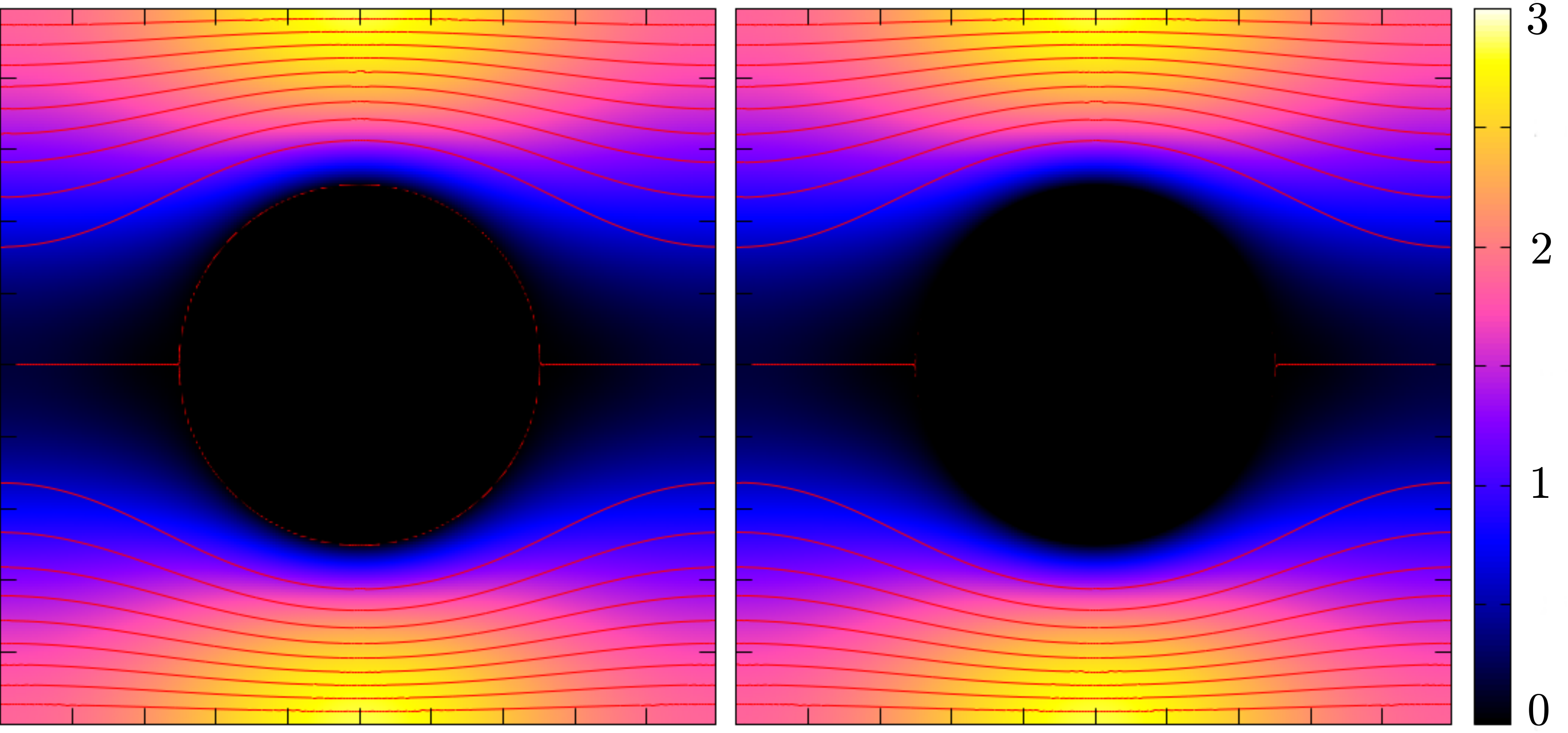}
  \caption{{\bf Upper:} Drag coefficient as a function of cylinder radius for Newtonian flow in the biperiodic geometry. Dotted line: analytical prediction in the limit $R\to 0$~\cite{Sangani1982}. Dashed line: analytical prediction in limit $2 R \to L_y$ of near-touching cylinders~\cite{Sangani1982}. Symbols: numerical results using immersed boundary method (circles), propagator method (squares), phase field method with $l=0.02$ (diamonds) and $l=0.01$ (triangles).  (See~\cite{HemingwayThesis2015} for definition of the parameter $l$.)\newline {\bf Lower:} Streamlines (red lines) and map of velocity magnitude $|\vtr{v}|$ calculated using IBM (left) and propagator (right) methods for a cylinder radius $R=0.25$ and overall throughput rate $Q=1.0$. }
\label{fig:Newtonian}
\end{figure}

The transfer of information between the Lagrangian and Eulerian grids
in Eqns.~\ref{eqn:porous_eul_to_lagr} and~\ref{eqn:porous_lagr_to_eul}
is achieved in numerical practice by approximating the Dirac delta
function $\delta(\vecv{r})$ by a smoothed Peskin delta function
$\delta_P(\vecv{r}) = \delta^x_P(x) \delta^y_P(y)$, in which
~\cite{Lai2000}
\begin{eqnarray}
8h  \delta_P(r) &=& 3 - 2\frac{|r|}{h} + \sqrt{+1 + 4\frac{|r|}{h} - 4\left(\frac{|r|}{h}\right)^2}  \;\; \;\;\; |r| \le h \nonumber \\
              &=& 5 - 2\frac{|r|}{h} - \sqrt{-7 + 12\frac{|r|}{h} - 4\left(\frac{|r|}{h}\right)^2}  \; h \le |r| \le 2 h \nonumber \\
              &=& 0 \;\;\;\;\;\; \;\;\;\;\;\; \;\;\;\;\;\; \;\;\;\;\;\; \;\;\;\;\;\;\;\;\;\;\; \;\;\;\;\;\; \;\;\;\;\;\; \;\;\;\;\;\;  \textrm{otherwise.}\nonumber
  \label{eqn:porous_peskin}
\end{eqnarray}
Here $h = \Delta x = L_x / N_x = \Delta y = L_y / N_y$ is the Eulerian
grid spacing, given a rectangular grid of $(N_x,N_y)$ points.  The
Lagrangian boundary of circumference $2\pi R$ is discretized into $M$
nodes of equal separation $\Delta s = 2\pi R / M$. The optimal value
of the ratio $\alpha = \Delta s / h$ is unknown {\it a priori}. Too small a
value risks oversampling the boundary forces, while too large a value
risks fluid leakage across the boundary.  We set $\alpha = 2$, and
have checked that our results are robust to reasonable variations
around this value.  We have also ensured that all our results
presented below are converged on the limit of grid spacing $h\to 0$
and of timestep $\Delta t\to 0$.

\begin{figure}[t]
  \includegraphics[width=8.0cm]{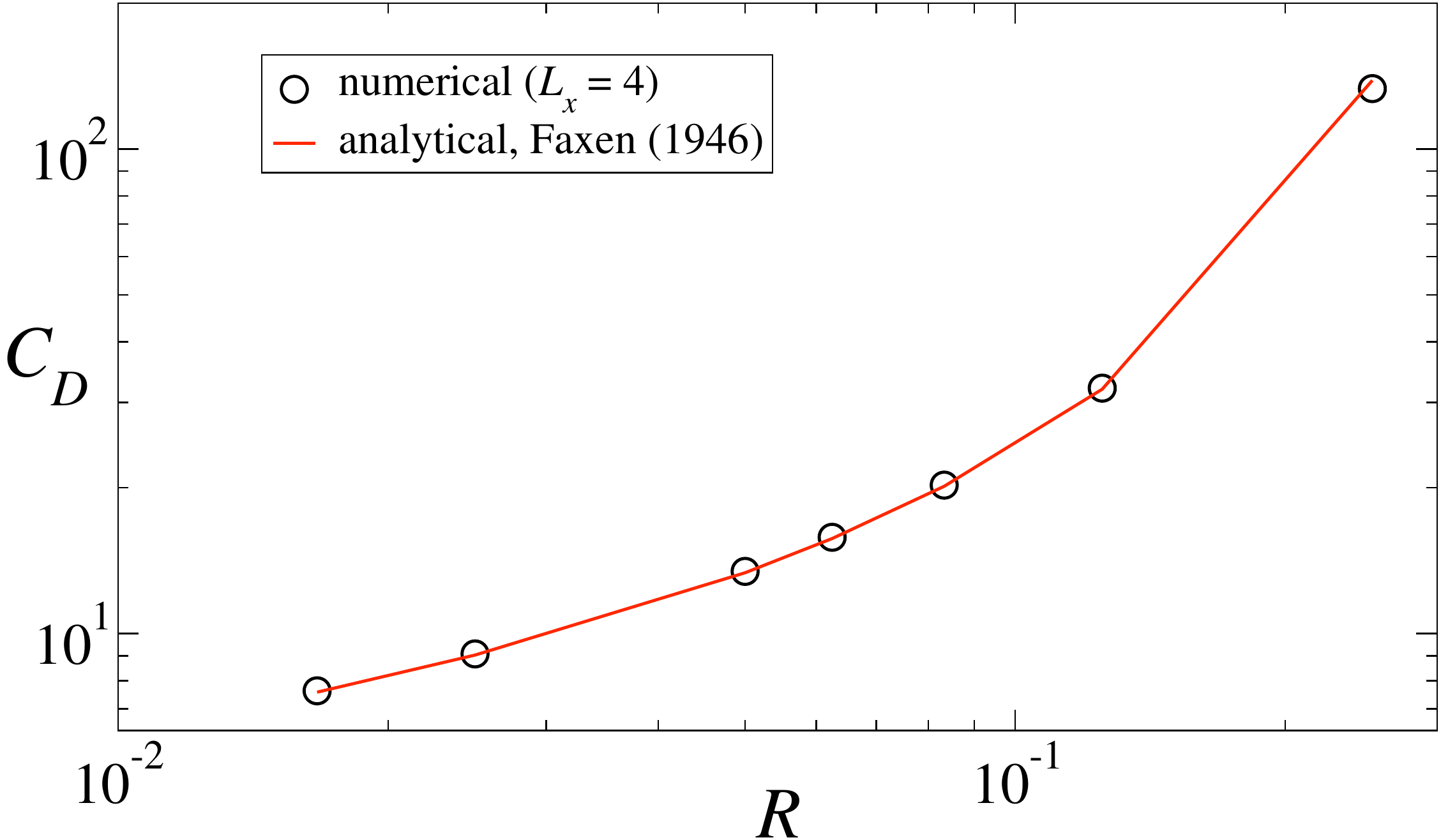}
  \vspace{5pt}

  \includegraphics[width=4.0cm]{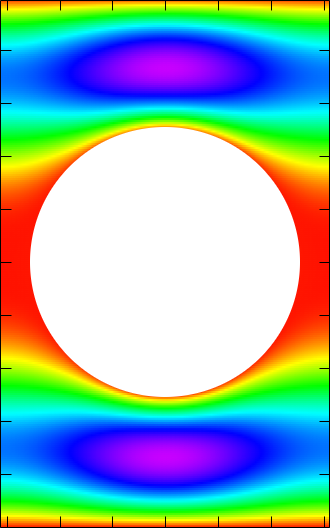}
  \includegraphics[width=4.0cm]{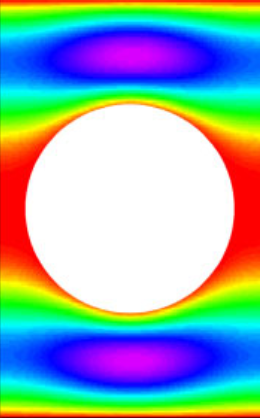}
  \caption{ {\bf Upper:} Comparison of the drag coefficient $C_D$ as obtained using IBM in the channel geometry for large cylinder spacing $L_x/L_y = 4$ with Faxen's analytical result for flow past a single cylinder in a channel~\cite{Faxens1946}.  {\bf Lower:} Map of velocity magnitude $|\vtr{v}|$ calculated using IBM (left) and from Ref.~\cite{Ellero2011} (right) for a cylinder radius $R=0.25$, cylinder spacing $L_x=0.625$ and throughput rate $Q=1.0$.  The colourscheme of this flow map is different from that of the others in this manuscript to match that of the original work~\cite{Ellero2011} against which we are benchmarking.  }
  \label{fig:Newtonian1}
\end{figure}

In the numerical solution just described, the methods employed to update the viscoelastic stress in the first substep of each timestep have been tested and benchmarked by ourselves in several previous publications~\cite{Fielding2006,Fielding2007}.  Therefore we focus here only on benchmarks to validate the second substep, in which the velocity field is calculated by solving Eqns.~\ref{eqn:Stokes} and~\ref{eqn:incomp}. The presence of viscoelasticity in this substep is trivial: it appears only as a source term $\nabla .\tens{\Sigma}$ in Eqn.~\ref{eqn:Stokes}). All the issues of principle are already contained in the solution of Eqns.~\ref{eqn:Stokes} and~\ref{eqn:incomp} for purely Newtonian flow in the geometries of interest here, for which known benchmarks exist, as follows.

Sangani and Acrivos~\cite{Sangani1982} derived analytical expressions
for the drag coefficient $C_D$ as a function of the cylinder radius
$R$ in Newtonian flow past the biperiodic array of
Fig.~\ref{fig:geometries} (upper), separately for the small cylinder
limit $R\to 0$, and for the limit $R\to L_y/2=1/2$ (in our units) in
which adjacent cylinders approach contact. These are shown by the
dotted and dashed lines respectively in the upper panel of
Fig.~\ref{fig:Newtonian}. Numerical results obtained using our IBM
method are in excellent agreement with these predictions, as shown by
the circles in the same panel.  Also shown (by the diamonds, squares
and triangles) are results for the drag coefficient additionally
obtained in the present study using two different numerical methods
that are independent of the IBM described above: a phase field method
and a circular-propagator method~\cite{HemingwayThesis2015}. As can be
seen, these numerical results for the drag coefficient are in
excellent agreement between all our three methods across the full
range of cylinder radii. For one particular value of the cylinder
radius, the full flow field as calculated in a biperiodic array is
shown in the lower panel of Fig.~\ref{fig:Newtonian}: on the left
using the IBM, and on the right using the propagator method.
Excellent agreement is seen between the two methods.

\begin{figure}[t]
  \centering
  \includegraphics[width=\columnwidth]{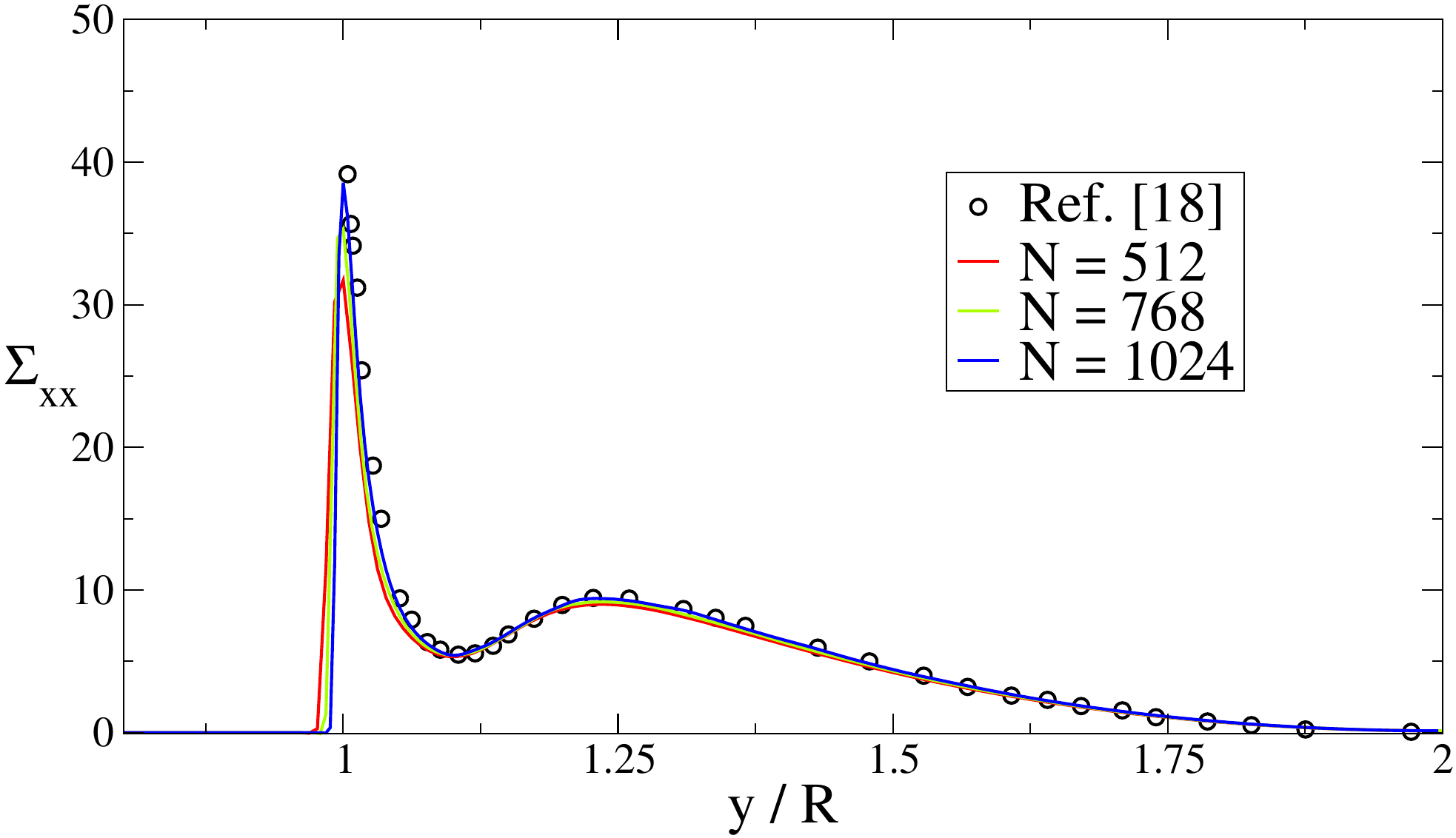}
  \caption{Comparison with data of Souvaliotis and Beris \cite{Souvaliotis1996} (black circles) for several grid resolutions $N$ (solid lines). The data were extracted from Ref.~\cite{Souvaliotis1996} using an online tool \cite{dataExtract}.  Shown are data taken along a vertical line through the cylinder centre, for the largest value of $\We^{\rm Beris} \equiv \tau V L_y / R^2 = 5.2$ explored in that study. $R / L_y = 0.25$, $\beta = 5/6$, $\ell = 0$. }
  \label{fig:beris}
\end{figure}

In Fig.~\ref{fig:Newtonian1} (upper), we show that numerical results
obtained using our IBM for the drag coefficient as a function of
cylinder radius $R$ for a fixed (wide) horizontal cylinder separation
$L_x=4.0$ in the channel geometry of Fig.~\ref{fig:geometries} (lower)
are in excellent agreement with the analytical prediction of
Faxen~\cite{Faxens1946} for flow past a single cylinder in a channel.
Fig.~\ref{fig:Newtonian1} (lower) shows that our results for the full
flow field in the channel geometry are in excellent agreement with
those of Ref.~\cite{Ellero2011}.  Finally, in Fig.~\ref{fig:beris} we
compare profiles of the polymer stress $\Sigma_{xx}$ against previous
work in the biperiodic geometry \cite{Souvaliotis1996}, again demonstrating excellent agreement.

\section{Results: viscoelastic flow past a biperiodic array of cylinders}
\label{sec:biperiodic}

Having carefully benchmarked our codes against known results in the literature, we now present our new results for viscoelastic flow past the biperiodic array of cylinders as sketched in Fig.~\ref{fig:geometries} (top).

\subsection{Character of the flow field}
\label{sec:character}

\begin{figure}[t]
  \centering
  \includegraphics[width=\columnwidth]{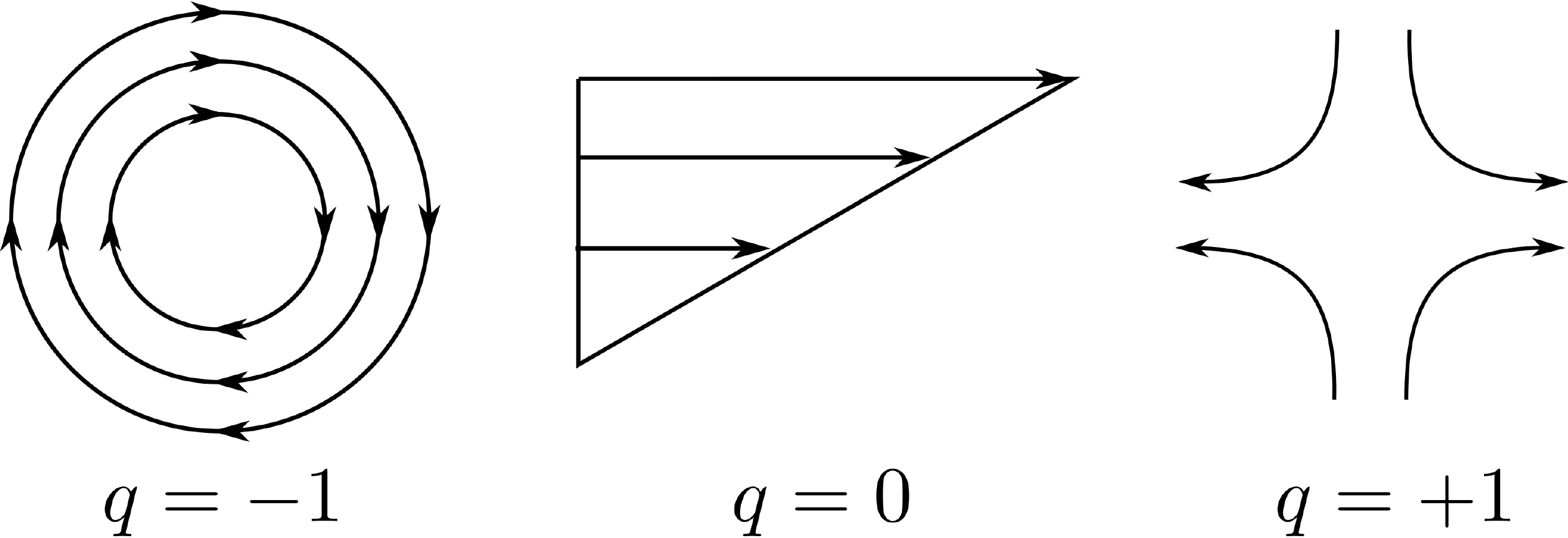}
  \caption{A schematic representation of the three flow types: {\bf
      left:} rotation, {\bf middle:} simple shear, and {\bf right:}
    extension (which is also called pure shear).}
  \label{fig:q_regimes}
\end{figure}

In Fig.~\ref{fig:constitutive} above we presented the stationary
constitutive curves of the Oldroyd B, FENE-P and FENE-CR models,
separately for the idealised flow fields of homogeneous simple shear
and homogeneous planar extension. A key aim of this work is to examine
whether the thickening observed as a function of increasing
Weissenberg number in the flow of a viscoelastic fluid through a
porous medium can be understood in terms of the fluid's underlying
constitutive behaviour in those simpler protocols of homogeneous
simple shear and planar extension.

In any porous flow geometry, however, the flow field will of course
vary in space (and sometimes also in time) in a complicated way, and
will in general comprise an admixture of both shear and extensional
components at any location. To quantify the flow field at any
location, therefore, we write the local velocity gradient tensor
$\tens{\gradv}$ as the sum of symmetric and antisymmetric contributions,
$\tens{D}=\tfrac{1}{2}(\tens{\gradv}+\tens{\gradv}^T)$ and
$\tens{\Omega}=\tfrac{1}{2}(\tens{\gradv}-\tens{\gradv}^T)$, with eigenvalues
$\lambdad=\sqrt{\tfrac{1}{2}\tens{D}:\tens{D}}$ and
$\lambdao=\sqrt{\tfrac{1}{2}\tens{\Omega}:\tens{\Omega}}$
respectively. Here $\lambdad$ measures the rate of deformation, and
accordingly the strength of the deforming effect that the flow is
expected to have on the polymer chains.  $\lambdao$ characterises the
rate of rotation, which does not deform the polymers.

Out of these two eigenvalues we also construct the frame-invariant,
rate-independent parameter~\cite{Skartsis1992,Gillissen2013}
\be
\label{eqn:nature}
q=\frac{\lambda_{\tens{D}}^2-\lambda_{\tens{\Omega}}^2}{\lambda_{\tens{D}}^2+\lambda_{\tens{\Omega}}^2}.
\ee
This quantifies the nature of the flow field at any location, in the
following way.  A value $q = +1$ corresponds to pure extensional flow,
which is sometimes also called pure shear flow (as expressed relative
to Cartesian axes in Eqn~\ref{eqn:ext}).  For $q = -1$ the flow is
purely rotational, and will have no deforming effect on the polymer
chains.  For $q = 0$ the rate of straining is equal to the rate of
rotation, giving an equal superposition of rotation and pure shear.
This corresponds to simple shear flow (as expressed relative to
Cartesian axes in Eqn.~\ref{eqn:shear}). These three cases $q=-1,0,+1$
are sketched in \figref{fig:q_regimes}.

\begin{figure*}[tbp]
  \includegraphics[width=1.0\textwidth]{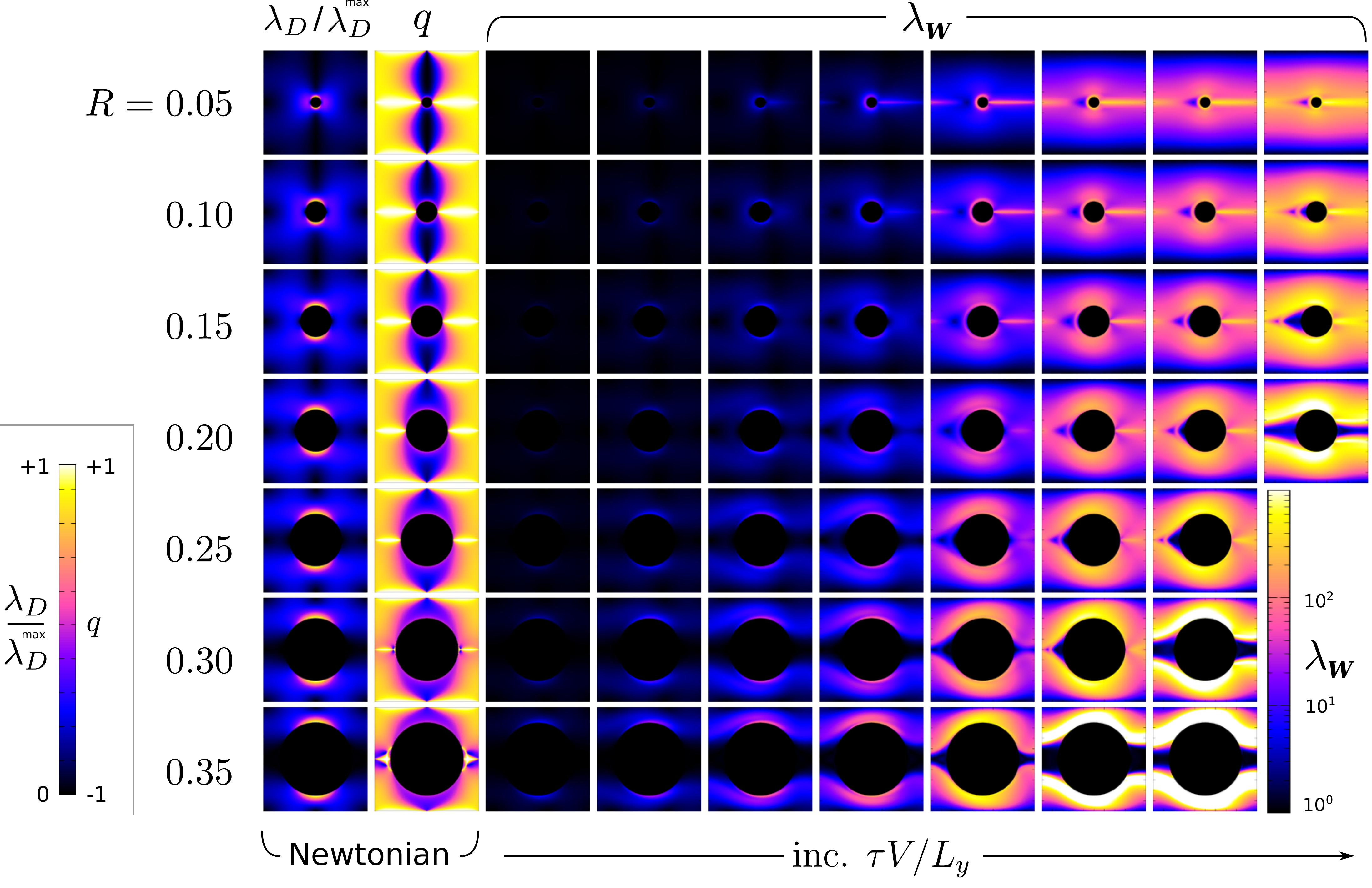}
  \caption{Montage of flow states for varying relaxation time and cylinder radius. Rows show cylinders with radii in the range $R = 0.05 \to 0.35$. The first column shows the maximum eigenvalue of the symmetrised velocity gradient tensor $\lambda_D$ (which we normalise to one), and the second column shows the flow character $q$ (colourbars are shown bottom left). The remaining columns show the principal eigenvalue $\lambdaconf$ of the polymer conformation tensor $\conf$ for $\tau V / L_y = 0.02, 0.05, 0.1, 0.2, 0.4, 0.8, 1.0, 1.6$ on a log colourscale (colourbar in bottom right). }
  \label{fig:montage}
\end{figure*}

\begin{figure}[t]
  \centering
  \includegraphics[width=\columnwidth]{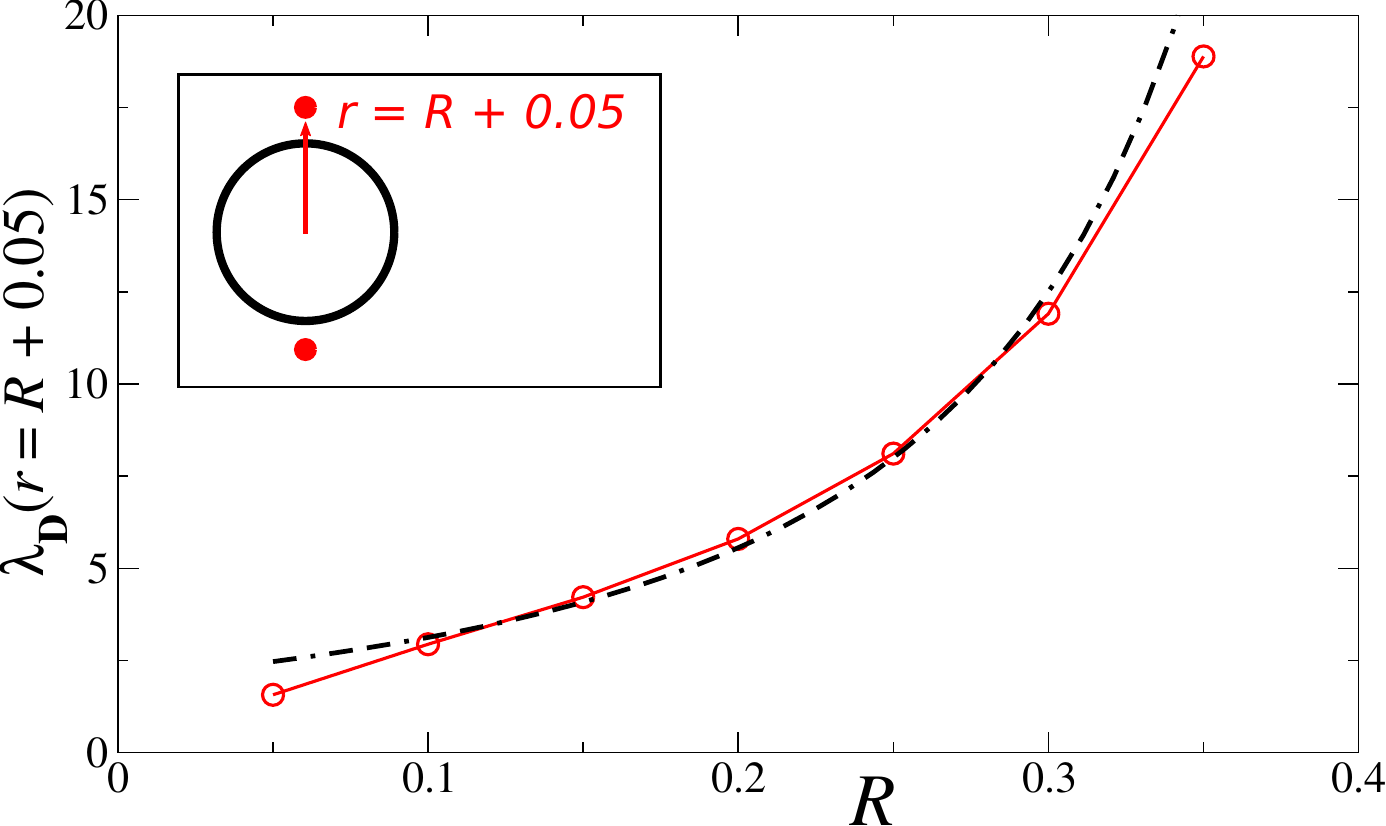}
  \caption{Scaling with cylinder radius $R$ of the rate of deformation $\lambda_{\tens{D}}$ in a Newtonian fluid (red solid line) in the regions just above and below the cylinder where the nature of the flow is simple shear. The plotted values of $\lambda_{\tens{D}}$ are taken at a distance $r = R + 0.05$ vertically from each cylinder centre (see red points in schematic in top left). The black dash-dotted line shows the scaling $2 V / \left( L - 2 R \right)^2$. }
  \label{fig:vert_vary_d}
\end{figure}

\begin{figure}[t]
  \centering
  \includegraphics[width=\columnwidth]{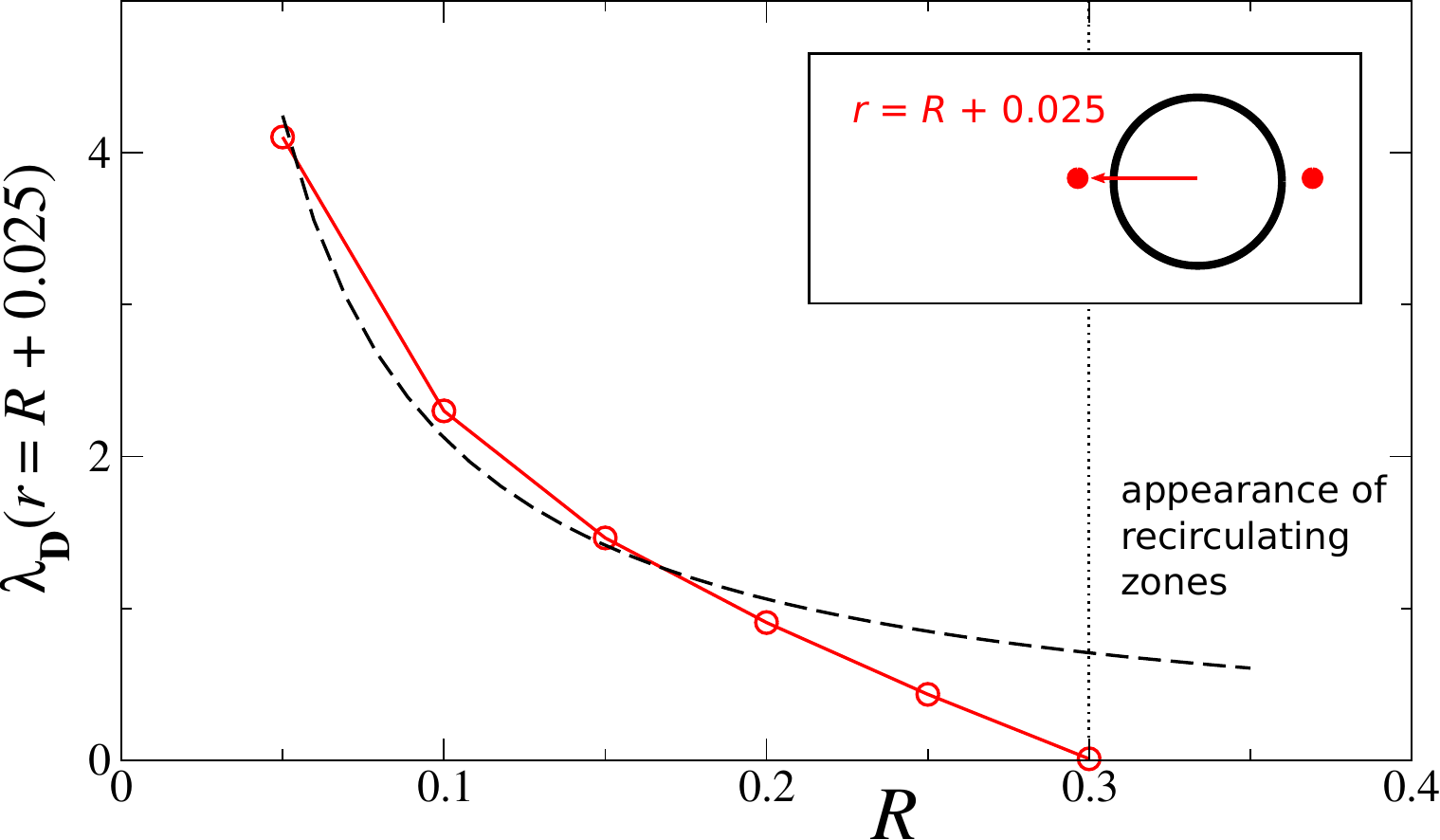}
  \caption{Scaling with cylinder radius $R$ of the rate deformation $\lambda_{\tens{D}}$ in a Newtonian fluid (red solid line) in the regions just fore and aft the cylinder where the nature of the flow is pure shear (\ie extension). The plotted values of $\lambda_{\tens{D}}$ are taken at a distance $r = R + 0.025$ horizontally from each cylinder centre (see red points in schematic in top right). The black dash-dotted line shows the scaling $\frac{2}{3} V/\left(\pi R\right)$. }
  \label{fig:hori_vary_d}
\end{figure}

\begin{figure}[t]
  \centering
  \includegraphics[width=\columnwidth]{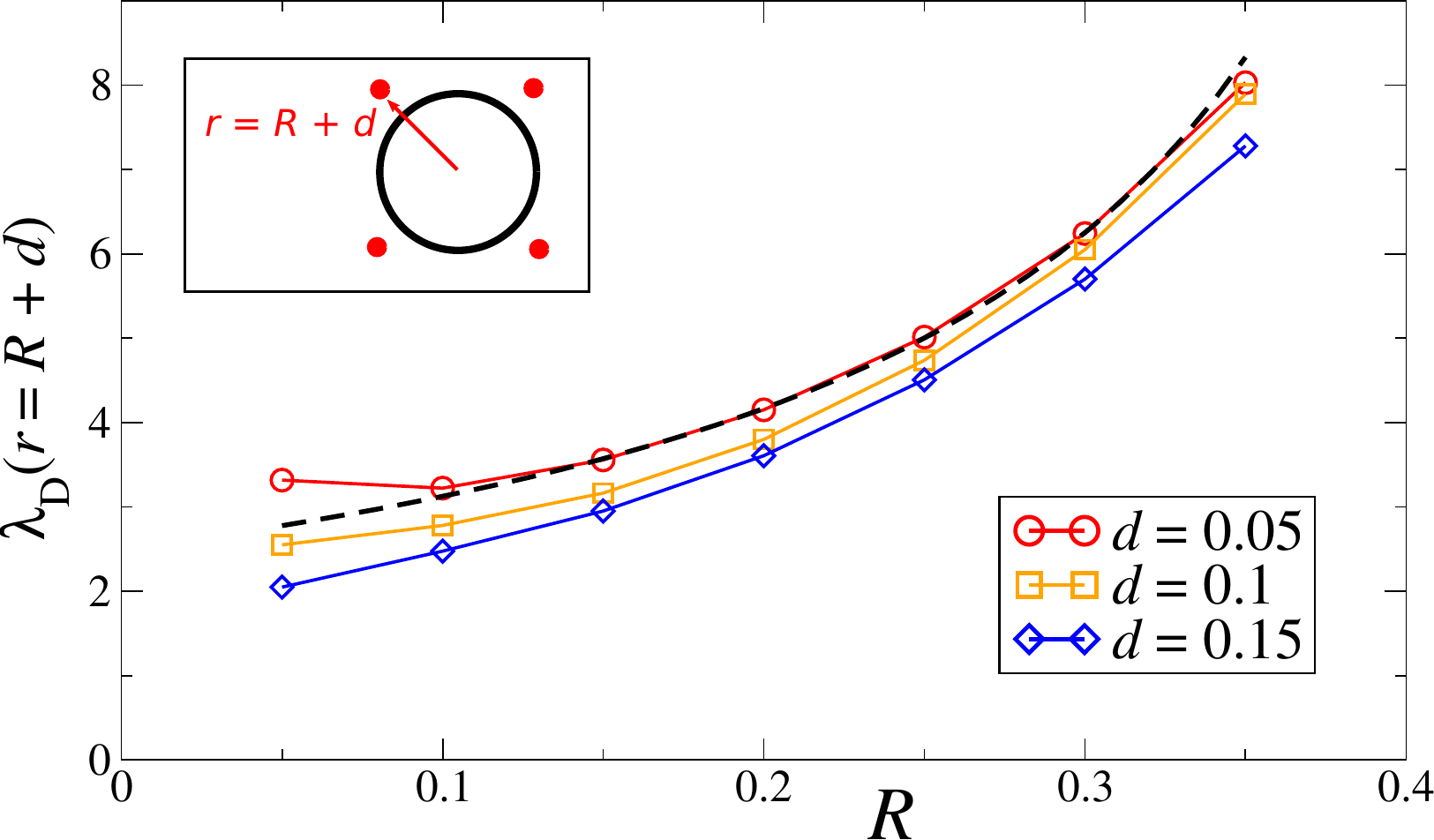}
  \caption{Scaling with cylinder radius of the rate of deformation $\lambda_{\tens{D}}$ in a Newtonian fluid in the regions centred along the diagonal $45^\circ$, \ie as the fluid just starts to squeeze into (and subsequently move out of) the contraction between vertically adjacent cylinders.  The plotted values of $\lambda_{\tens{D}}$ are taken at distances $r = R + d$ along any diagonal angled at $45^{\circ}$ from the cylinder centre (see red points in schematic in top left). All four points are equivalent due to the symmetry of the Newtonian flow field.  The black dash-dotted line shows the scaling $2.5 V / \left(L - 2 R \right)$. }
  \label{fig:diag_vary_d}
\end{figure}

In practice, of course, the flow field in any given geometry will change as a function of Weissenberg number $\We$.  Indeed, this effect is fully accounted for in our numerical studies, as described in Sec.~\ref{sec:numerics} above. In examining our numerical results, however, we find that this change is relatively modest.  Because of this, we shall frame the discussion of our results in trying to understand how the polymer responds, with increasing $\We$, to a velocity field that is assumed to be unchanged from that at $\We=0$.  Accordingly, we focus the discussion in this section on the flow field that pertains at $\We=0$.

As can be seen in the left two columns of Fig.~\ref{fig:montage}, the rate of deformation $\lambdad$ is strongest round the upper and lower edges of the cylinder for all values of the cylinder radius $R$, and is dominated by simple shear.  As shown in Fig.~\ref{fig:vert_vary_d}, the deformation rate in these regions scales with cylinder radius $R$ as $V L/(L - 2R)^2$.  Because the flux is constant through any vertical slice, the mean velocity  $V$ increases as it is forced through the narrow vertical gap between adjacent cylinders, producing an effective velocity that might be expected to scale as $V_{e} \sim V L / (L - 2R)$. A typical shear-rate in the gap would then be $V_{e} / (L - 2R) = V L / (L - 2R)^2$.  Based on this, we define a Weissenberg number
\be
\Wezero=\frac{VL\tau}{(L-2R)^2}.
\label{eqn:wezero}
\ee
If any regime exists in which the pressure drop is dominated by these
regions of simple shear just above and below the cylinder, we would
expect $\Wezero$ to be the relevant Weissenberg number to characterise
that regime.

For the low porosity geometries with small $R$, there also exists a
region of reasonably strong deformation fore and aft of the cylinder,
which is extensionally dominated.  To characterise this, we define a
Weissenberg number
\be
\Weone=\frac{V\tau}{\pi R}.
\label{eqn:weone}
\ee
This takes as its characteristic time the residence time of the polymer near the cylinder, $\pi R / V$.   The cylinders are widely spaced in this regime, so we expect the inter-cylinder spacing to be a less important lengthscale in comparison. As shown in \figref{fig:hori_vary_d}, this definition provides a reasonable approximation of the deformation rate in this region. Therefore in any regime in which the pressure drop is dominated by these regions of extension fore or aft of the cylinder, we expect $\Weone$ to be the relevant Weissenberg number to characterise the flow in that regime. Note that while $\Weone$ should strictly be labelled as a Deborah number \cite{Colby2013}, for simplicity here we retain the label $\Weone$.

Finally, there exists a region of moderately strong extensional flow centred around the 45$^\circ$ diagonal lines, associated with the fluid just starting to squeeze into, and then subsequently move out of, the gap between vertically adjacent cylinders. As shown in Fig.~\ref{fig:diag_vary_d}, the deformation rate in this region scales as $V/(L-2R)$. Accordingly, we define a Weissenberg number
\be
\Wetwo=\frac{V\tau}{(L-2R)}.
\label{eqn:wetwo}
\ee
In any regime where the pressure drop is dominated by the squeezing of the fluid through the gap between vertically adjacent cylinders, we expect $\Wetwo$ to be the relevant Weissenberg number to characterise the flow. Note that while the scalings provided by \figsref{fig:hori_vary_d}-\ref{fig:diag_vary_d} are taken at representative points in the fluid, they can only expected to hold to within an order 1 prefactor. However we will show that each of the Weissenberg number definitions \eqsref{eqn:weone},\ref{eqn:wetwo} accurately captures the onset of thickening in the regime of $R$ for which it is expected to apply, justifying our choices.

Having discussed the nature of the flow field in the limit of Newtonian flow, $\We\to 0$, we now proceed to describe our numerical results for the flow response of the Oldroyd B, FENE-P and FENE-CR models as the Weissenberg number $\We$ (according, in any regime, to the most relevant of the above definitions in that regime) increases with the polymer relaxation time $\tau$.

\subsection{Oldroyd B model}

\begin{figure}[t]
  \centering
  \includegraphics[width=0.5\textwidth]{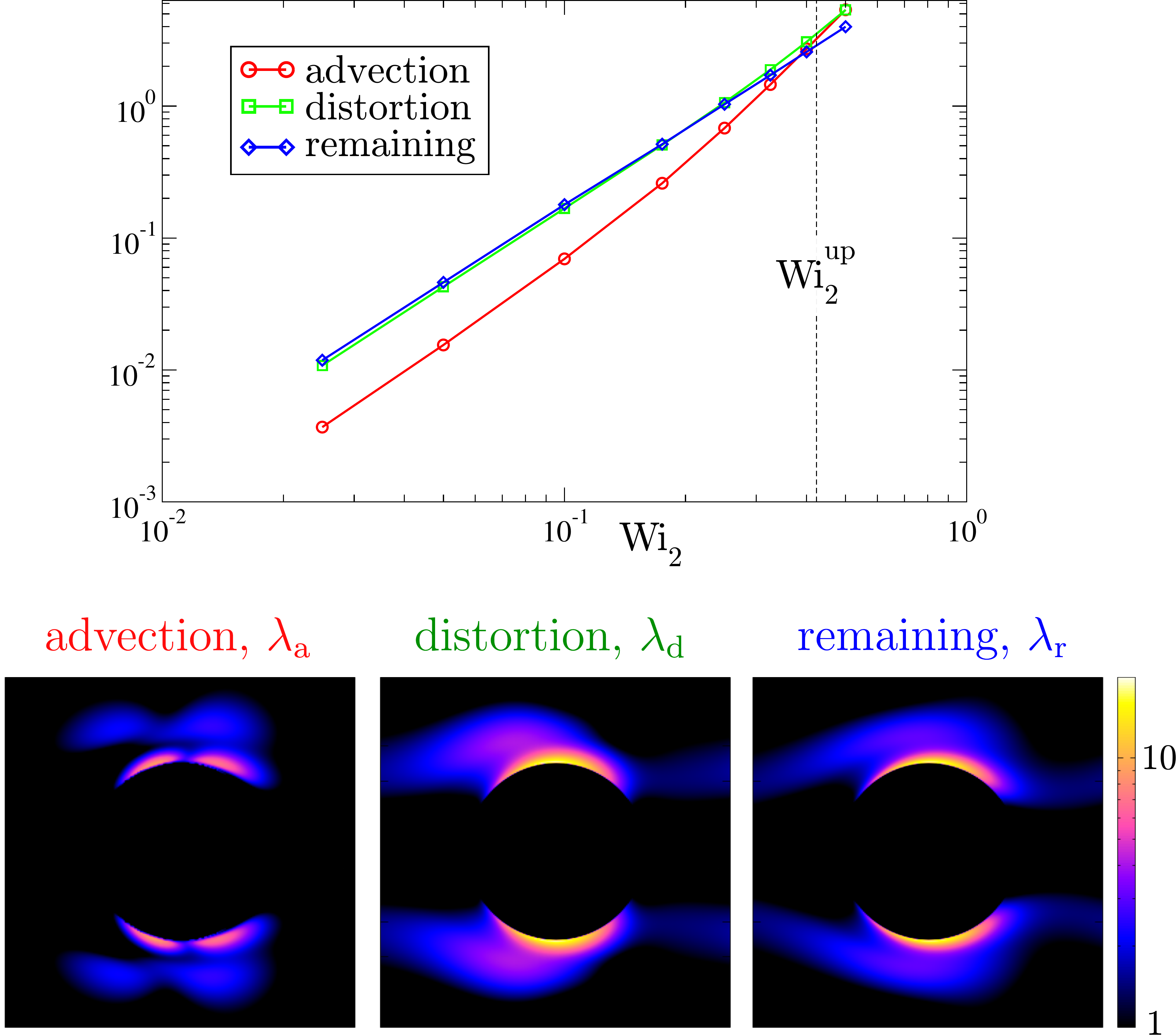}
  \caption{\textbf{Top:} Integrals of $\lambda_\textrm{a}, \lambda_{\rm d}, \lambda_{\rm r}$ (see \eqref{eq:contract}) over space for several values of $\We_2$. Also marked is the point of upturn $\We_2^{\rm up}$ (dashed line), defined as the Weissenberg number for which the minimum drag occurs. \textbf{Bottom:} Colourmaps of $\lambda_\textrm{a}, \lambda_{\rm d}, \lambda_{\rm r}$ for $\We_2 = 0.25$. Note that in these runs only we have removed the diffusive contribution to the dynamics by setting $\ell = 0$.  $R / L_y = 0.25$. }
  \label{fig:convect_distort}
\end{figure}

\begin{figure}[t]
  \centering
  \includegraphics[width=0.9\columnwidth]{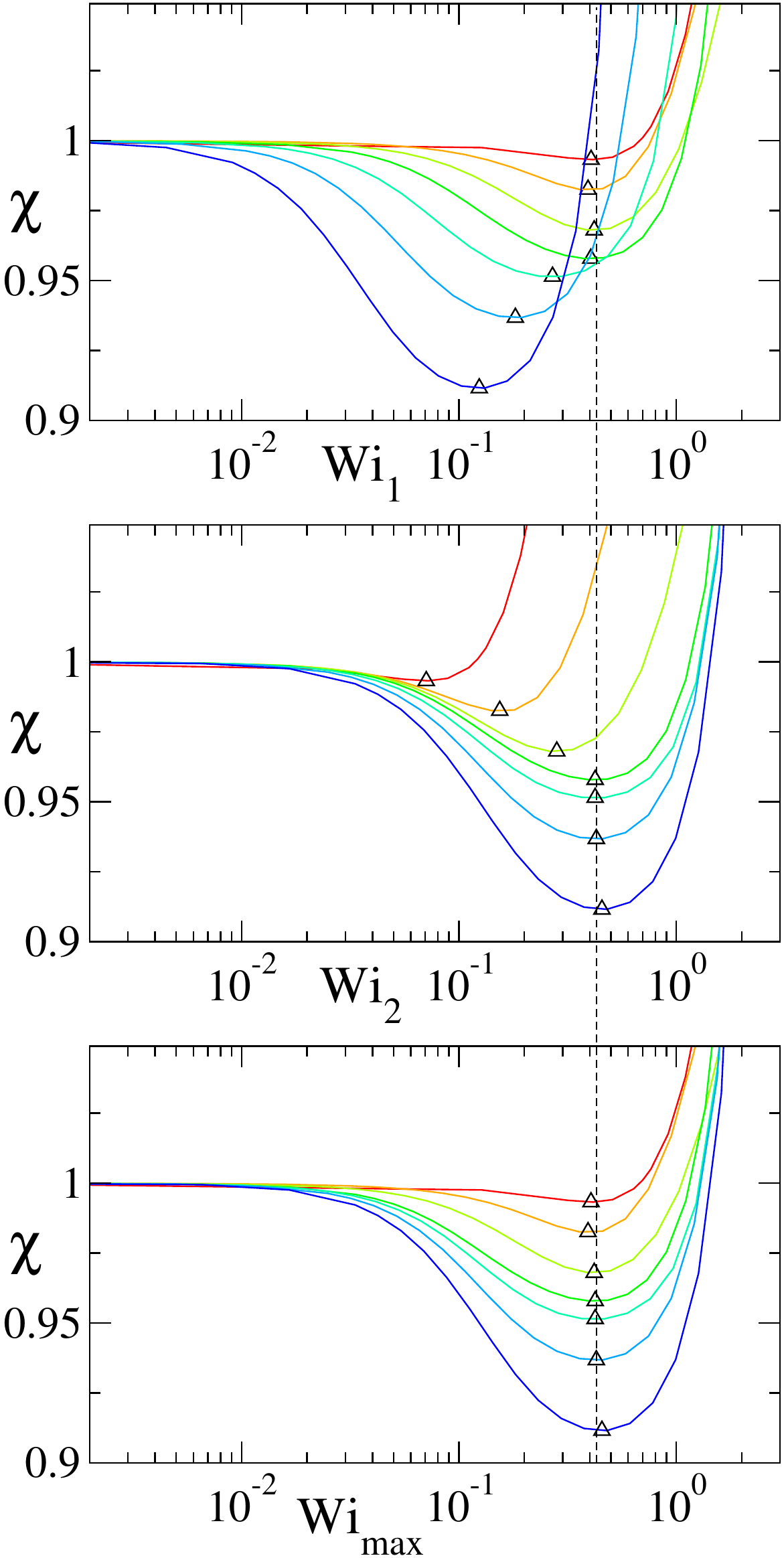}
  \caption{Normalised drag coefficient $\chi$ for the Oldroyd B model in the biperiodic geometry of Fig.~\ref{fig:geometries} (upper) plotted against (a) the Weissenberg number $\Weone$ of Eqn.~\ref{eqn:weone}, (b) the Weissenberg number $\Wetwo$ of Eqn.~\ref{eqn:wetwo} and (c) the maximum of these two Weissenberg numbers.  Curves downwards at $\We=0.1$ in each panel correspond to cylinder radii $R = 0.05$ (red) $\to 0.35$ (blue) in increments of $\Delta R = 0.05$. Triangles mark the point of upturn, defined as the minimum of $\chi(\We)$.}
  \label{fig:drag}
\end{figure}

In Fig.~\ref{fig:montage} we explore the flow response of the Oldroyd
B model as a function of the cylinder radius $R$ (down each column)
and the adimensional polymer relaxation time $\tau$, proportional to
the Weissenberg number (along each row).  (In each case the cell
height $L_y=L_x\equiv L=1.0$ and the velocity characterising the
throughput $V=1.0$.)

Shown in each column (beyond the first two) of Fig.~\ref{fig:montage} is a colourmap of the largest eigenvalue $\lambdaconf$ of the polymer conformation tensor $\conf$, which characterises the degree to which the polymer molecules are deformed by the flow.  ($\conf=\tens{\delta}$ in a fluid equilibrated at rest.) For small Weissenberg number (third and fourth columns), the colourmap of the polymer deformation $\lambdaconf$ is essentially the same as that of the deformation rate of the flow field $\lambdad$ (first column), consistent with the fact that the solution of the Oldroyd B model in the Newtonian limit $\tau\to 0$ is $\conf=\tens{\delta} + 2\tau\tens{D}$.

Moving rightwards across the montage from the third column, $\tau$
(and so $\We$) progressively increases. For each cylinder radius $R$
(along each row), the first noticeable effect with increasing $\tau$
is that the bright regions of strong polymer deformation associated
with the regions of simple shear along the top and bottom of the
cylinder shift slightly downstream (rightwards).  Given that this
shift is in the direction of flow, one might expect
that this arises due to the increasing influence of the advective
term $\vecv{v}.\nabla\conf$ in the polymer constitutive equation with $\tau$.
(In contrast, for $\tau=0$ the advection term plays no role and the
dynamics are purely local.) The polymer conformation at any location
would then be affected not only by the flow field at that location, but
would also receive information about the flow field immediately upstream
(leftwards).

However, elastic effects arising due to the distortion term cannot
be neglected. In order to quantify the relative strength of advection
and distortion, we first note that the constitutive equation \eqref{eqn:Maxwell} can be rewritten in
terms of the polymer stress (as opposed to the conformation tensor $\conf$). In steady
state this reads
\begin{equation}
  \underbrace{\left(\vtr{v} \cdot \nabla \right) \tsr{\Sigma}}_{\textrm{advection, } \tsr{T}_\textrm{a} } = \underbrace{\tsr{\gradv}^T\cdot\tsr{\Sigma} + \tsr{\Sigma}\cdot\tsr{\gradv}}_{\textrm{distortion, } \tsr{T}_\textrm{d}} + \underbrace{2 G \tsr{D} - \frac{1}{\tau} \tsr{\Sigma}}_{\textrm{remaining, } \tsr{T}_\textrm{r}}.
\end{equation}
The three tensorial contributions can be contracted to scalars as
\begin{align}
  \lambda_{\rm a} = \sqrt{\tsr{T}_a : \tsr{T}_a}, \;
  \lambda_{\rm d} = \sqrt{\tsr{T}_d : \tsr{T}_d}, \;
  \lambda_{\rm r} = \sqrt{\tsr{T}_r : \tsr{T}_r},
  \label{eq:contract}
\end{align}
and integrated over space to yield a single value for each term.  In Fig.~\ref{fig:convect_distort} (top) we plot these integrals as a function of $\We_2$. This shows that for small to moderate values of $\We_2$, the steady-state equation is dominated by distortion rather than advection. This result is reinforced by the colourmaps in Fig.~\ref{fig:convect_distort} (bottom) that show that distortion (near the cylinder tops) is stronger than advection (just before and after the narrowest vertical point).  This suggests that the observed downturn in the drag is mainly a result of distortion rather than advection. Interestingly, beyond a certain value of $\We_2$, the advective term becomes comparable to distortion. We will later show that this coincides with the point at which the drag dramatically increases, $\We_2^{\rm up}$ (see black triangles in \figref{fig:drag}).

Corresponding to the montage of Fig.~\ref{fig:montage}, curves of the
normalised drag coefficient $\chi$ as a function of increasing $\We$
are shown in Fig.~\ref{fig:drag}, separately for each value of $R$
(each row of the montage). The signature in the drag coefficient of
the initial slight shift rightwards of the flow pattern just described
is an initial decrease of $\chi$ with $\We$, relative to the value
$\chi=1.0$ in the Newtonian limit $\We\to 0$.  In the existing
literature this effect is sometime suggested to stem from a shear
thinning effect~\cite{Moss2010,Gillissen2013}.  However the Oldroyd B
model does not shear thin, so that explanation cannot be valid here.
We suggest instead that this initial decrease in drag arises due to
the effect of the elastic distortion terms in slightly (spatially) `delaying'
the build-up of viscoelastic stress at any location (recall Fig. \ref{fig:convect_distort}),
relative to the Newtonian case. In this way, the regions of strongest polymer
deformation are shifted slightly away from the regions of strongest
shear rate.

\begin{figure}[t]
  \includegraphics[width=\columnwidth]{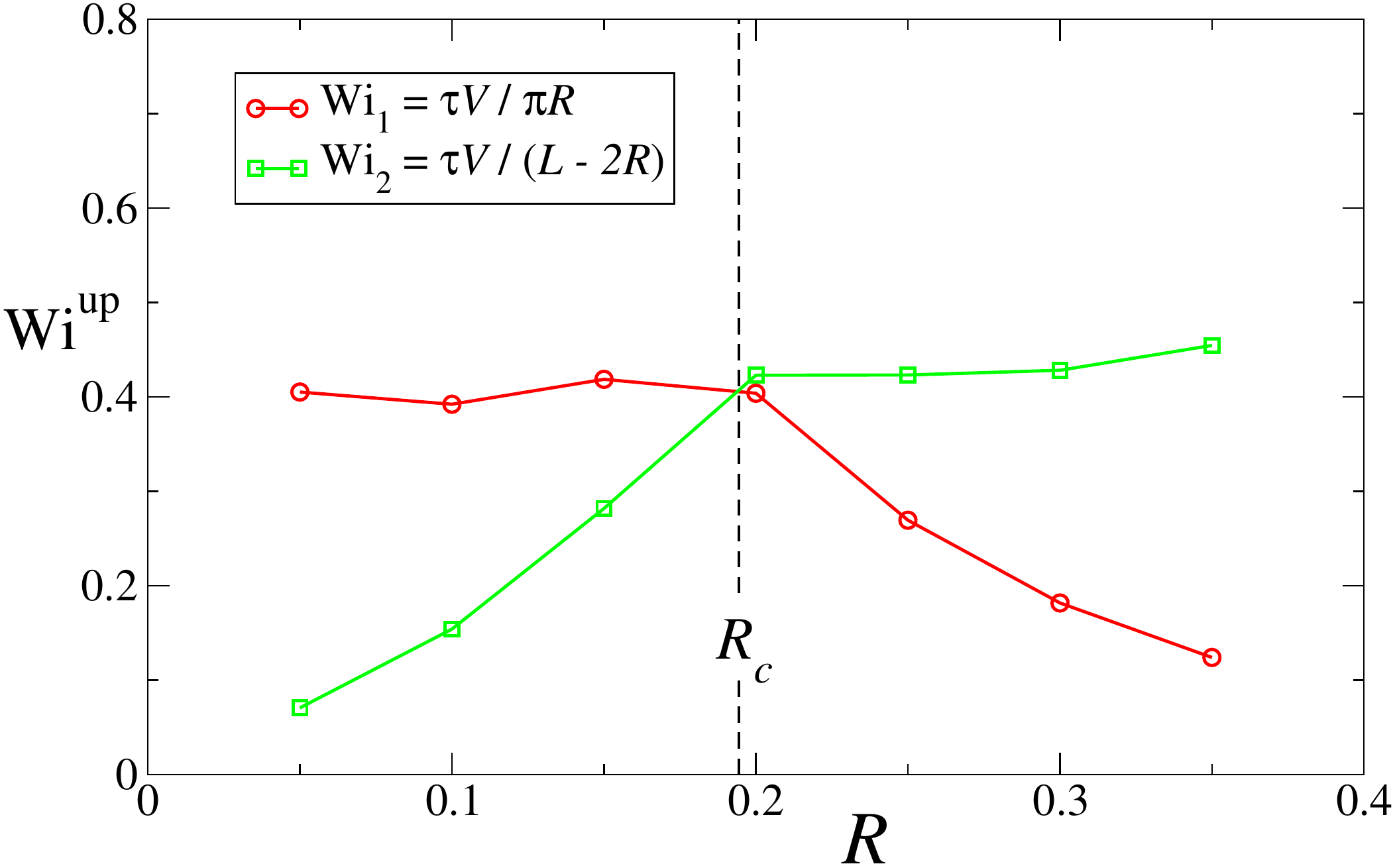}
  \caption{Dependence on cylinder radius $R$ of the value of the Weissenberg number $\We=\Weup$ at which the upturn in the drag coefficient occurs.  For large $R$, $\Wetwo$ captures a single value for the upturn, whereas for small $R$ (where extension dominates), $\Weone$ captures a single value.}
\label{fig:crossover}
\end{figure}

Moving further right across the montage of Fig.~\ref{fig:montage}, more dramatic changes in the flow field are evident.  For small values of the cylinder radius $R$, as $\tau$ (and so $\We$)
increases rightwards across the montage, a bright streak develops in
the colourmap of the polymer conformation tensor, focused in a wake
along the centreline aft of the cylinder. This is due to the region of
extensional flow to the right of the stagnation point at the
centre-aft point of the cylinder edge, which causes a strong
extensional stretching of the polymer molecules as the Weissenberg
number increases.  (Recall this region of extensional flow, $q=1$, in
the first two columns of the montage for small $R$. As noted above, the
flow map itself also changes with $\We$, but to a relatively modest
extent.)  This extensional stretching of the polymer chains manifests
itself as a strong upturn in the drag coefficient following a minimum
at $\Weup$.  As can be seen in Fig.~\ref{fig:drag}a), this occurs at a
(nearly) constant value of the Weissenberg number $\Weone$ for several
different (small) values of the cylinder radius $R$. This confirms
that $\Weone$, which we recall is defined by considering the rate of
extensional deformation aft (and fore) of each cylinder, is indeed the
relevant dimensionless rate to characterise the flow in this regime of
low cylinder radius and high medium porosity.

James studied experimentally the flow of a Boger fluid past a square array of cylinders in this regime of small $R/L = 0.09 \to 0.18$~\cite{James2012}.  As in our simulations, they found the flow to be steady (up to values of their defined Deborah number $De \equiv \tau V/L\approx 4$). They likewise reported only a small downturn in the drag coefficient as a function of increasing $De$, before a pronounced upturn at $De = 0.5$.

Returning to \figref{fig:montage}, for larger values of the cylinder radius $R$, the dominant effect as
$\tau$ increases rightwards across the montage is that the layers of
strong polymer deformation in the shear fields on the upper and lower
edges of the cylinder intensify, and are supplemented by the
development of secondary `layers' of strong polymer deformation above
and below the cylinder (one layer above the cylinder and one below
it). These appear to originate in the contraction as the flow moves
from the left hand edge of each snapshot into the narrow gaps between
vertically adjacent cylinders, with these secondary `layers' also
advected downstream into the vertical gap.  This again manifests itself as
a strong upturn in the drag coefficient following a minimum at
$\Weup$.  As seen in Fig.~\ref{fig:drag}b), this upturn occurs at a
constant value of $\Wetwo$ for several different values of the
cylinder radius $R$. This confirms that $\Wetwo$ is indeed the
relevant Weissenberg number to characterise the flow in this regime of
larger cylinder radius and smaller medium porosity.

The values of $\Weone$ and $\Wetwo$ at the minima in the drag
coefficient curves of Fig.~\ref{fig:drag} are plotted as a function of
cylinder radius $R$ in Fig.~\ref{fig:crossover}. As can be seen, the
upturn occurs at a roughly constant value of
\be
\Wemax={\rm  max}(\Weone,\Wetwo).
\label{eqn:wemax}
\ee
The crossover between which of $\Weone$ and $\Wetwo$ dominates at any
value of $R$ occurs at $\Rcross=L/(2+\pi)\approx 0.194$. For cylinder
radii $R<\Rcross$ we have $\Weone>\Wetwo$: in this regime, the effects
of the extensional wake in the relatively wide horizontal gap between
horizontally adjacent cylinders dominate those of shear in the gap
between vertically adjacent cylinders. For $R> \Rcross$ we have
$\Wetwo>\Weone$: in this regime the squeezing of the fluid into the
now narrower vertical gap between vertically adjacent cylinders
dominates any effects in the now smaller horizontal extensional wake.
This effectiveness of $\Wemax$ in characterising the flow across the
full range of cylinder radii is also seen via the master plot of the
drag coefficient $\chi$ as a function of $\Wemax$ in
Fig.~\ref{fig:drag}c): the upturn occurs at a fixed value of $\Wemax$
for all values of $R$.

\begin{figure}[t]
  \includegraphics[width=0.48\textwidth]{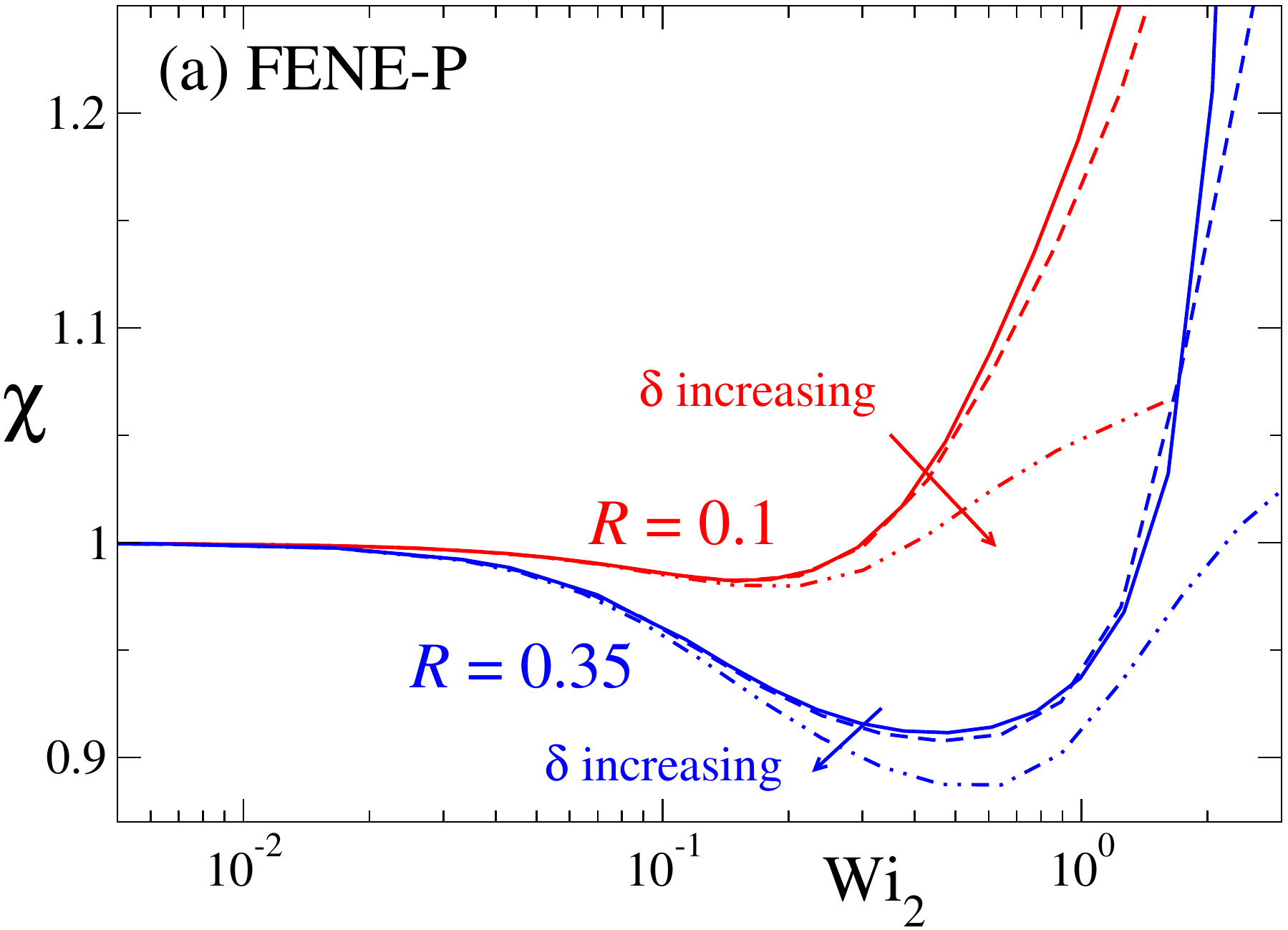}
  \includegraphics[width=0.48\textwidth]{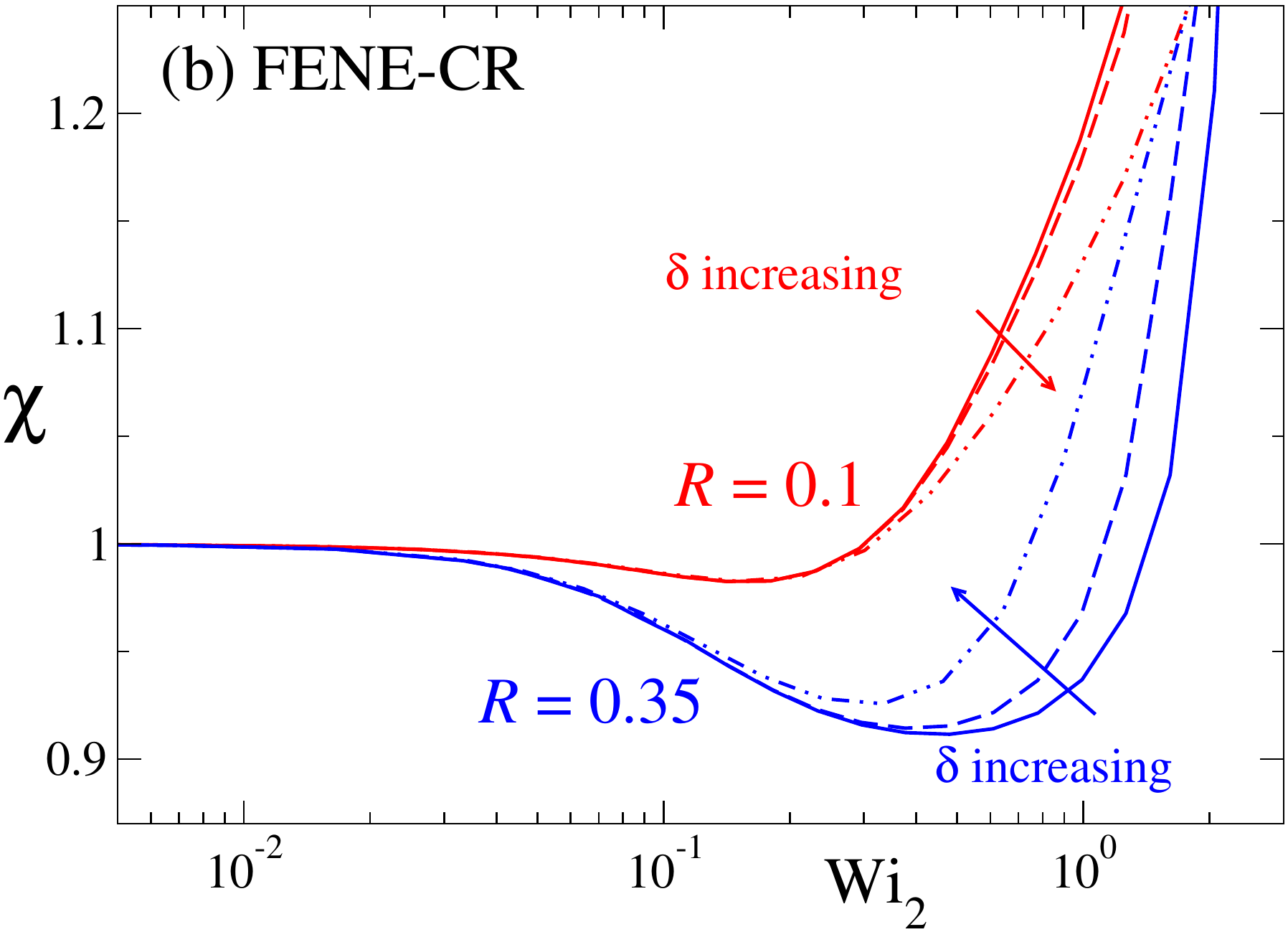}\hfill
  \caption{Drag coefficient as a function of Weissenberg number $\Wetwo$ in the (a) FENE-P and (b) FENE-CR models for $\delta=0.0$ (solid lines, recovering Oldroyd B behaviour), $\delta=0.001$ (dashed lines) and $\delta=0.01$ (dot-dashed lines). In each panel data are shown for a cylinder radius $R=0.1<\Rcross$ (red curves) and a cylinder
  radius $R=0.35>\Rcross$ (blue curves). }
  \label{fig:fene}
\end{figure}

\subsection{Fene models}

We consider now the effect of finite dumbbell extensibility on the phenomena just discussed. In particular, we shall report the drag coefficient as a function of Weissenberg number in each of the FENE-P and FENE-CR models, for two different values of the cylinder radius: $R=0.1<\Rcross\approx 0.194$ and $R=0.35>\Rcross\approx 0.194$.  The case $R=1/2.4\approx 0.41$ was considered previously~\cite{LiuThesis1997} for both FENE models, and our findings for $R=0.35$ will be qualitatively consistent with that study.

Our results for the FENE-P model are shown in Fig.~\ref{fig:fene}a).  We recall from Fig.~\ref{fig:constitutive} that, under conditions of homogeneous viscometric flow, this model thins in both shear and extension.  Consistently, we find that the drag coefficient is smaller for the FENE-P model ($\delta>0$) than for the Oldroyd B model ($\delta=0$) for both values of $R$ considered.

Our results for the FENE-CR model are shown in Fig.~\ref{fig:fene}b).  We recall from Fig.~\ref{fig:constitutive} that, under conditions of homogeneous viscometric flow, this model thins only in extension but not in shear. In the porous geometry studied here, for a cylinder radius $R=0.1<\Rcross$ the drag coefficient is lower in FENE-CR ($\delta>0$) than in Oldroyd B ($\delta=0$).  This is consistent with the extensional thinning of FENE-CR, and with the fact that the flow is dominated by the extensional wake aft the cylinder for this value of $R$. In contrast, for a cylinder radius $R=0.35>\Rcross$, the drag coefficient is {\em larger} in FENE-CR than in Oldroyd B. Clearly, this observation lacks any obvious explanation in terms of the homogeneous constitutive curves of Fig.~\ref{fig:constitutive}.  Feasibly, it could arise because the finite dumbbell extensibility reduces the extent to which the molecules are stretched and reoriented as they transit the contraction flow en route into the gap between vertically adjacent cylinders, causing them then to confer a greater shear stress in that gap. We do not provide any evidence to support this claim, however.

\section{Results: viscoelastic flow past an array of cylinders in a channel}
\label{sec:channel}

\begin{figure*}[t]
  \includegraphics[width=\textwidth]{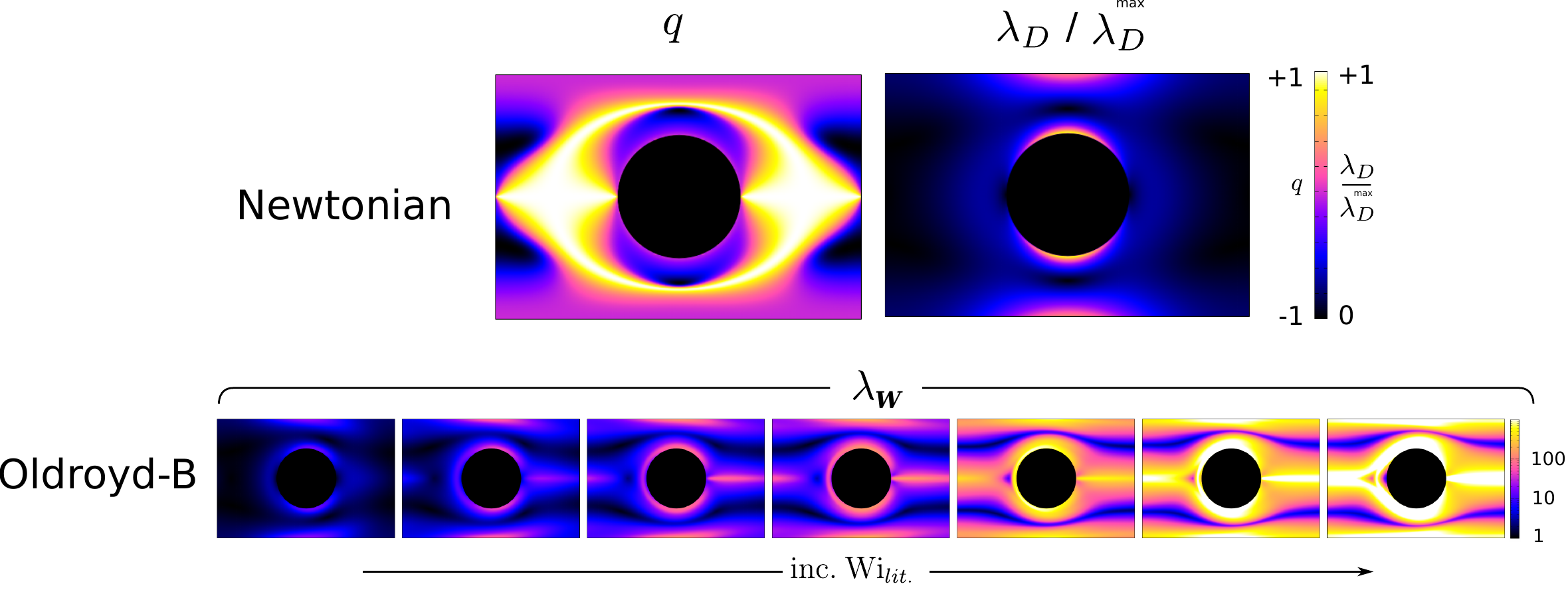}
  \caption{{\bf Upper:} flow character $q$ and deformation rate $\lambda_D$ for Newtonian flow past a periodic array of widely spaced cylinders with $L_x/L_y=1.5$, $R/L_y = 0.25$. {\bf Lower:} polymer deformation $\lambda_{\bf W}$ for the Oldroyd-B model as a function of of increasing Weissenberg number ($\We_{lit.}=\tau V / R = 0.31, 0.64, 0.95, 1.24, 2.41, 3.28, 3.95$ from left to right) in the same geometry.}
  \label{fig:porous_Qc_grid}
\end{figure*}

\begin{figure*}[t]

  \includegraphics[width=\textwidth]{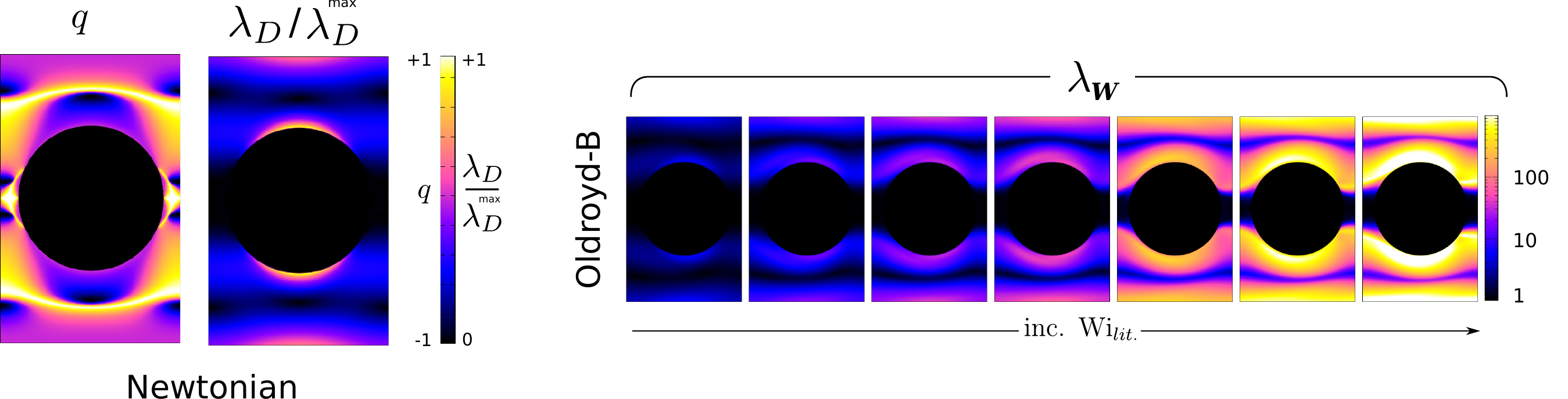}
  \caption{{\bf Left:} flow character $q$ and deformation rate $\lambda_D$ for Newtonian flow past a periodic array of closely spaced cylinders with $L_x/L_y=0.625$, $R/L_y=0.25$.  Right: polymer deformation $\lambda_{\bf W}$ for Oldroyd-B model as a function of increasing Weissenberg number ($\We_{lit.} = \tau V / R = 0.0009, 0.009, 0.074, 0.30, 0.60, 1.41, 2.61$ from left to right) in the same geometry. }
  \label{fig:porous_Qd_grid}
\end{figure*}

\begin{figure*}[t]

  \includegraphics[width=0.94\columnwidth]{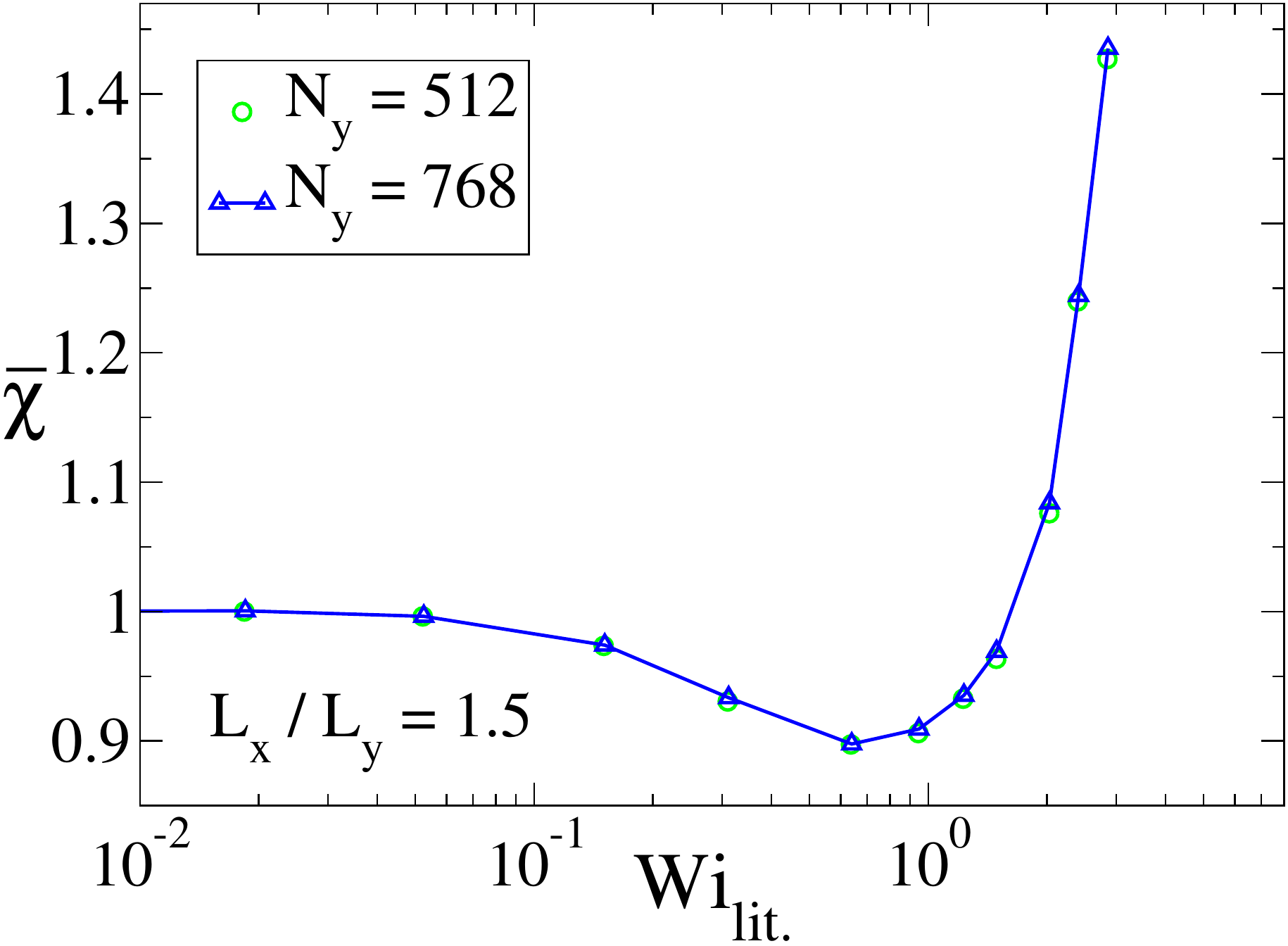}
  \hfill
  \includegraphics[width=\columnwidth]{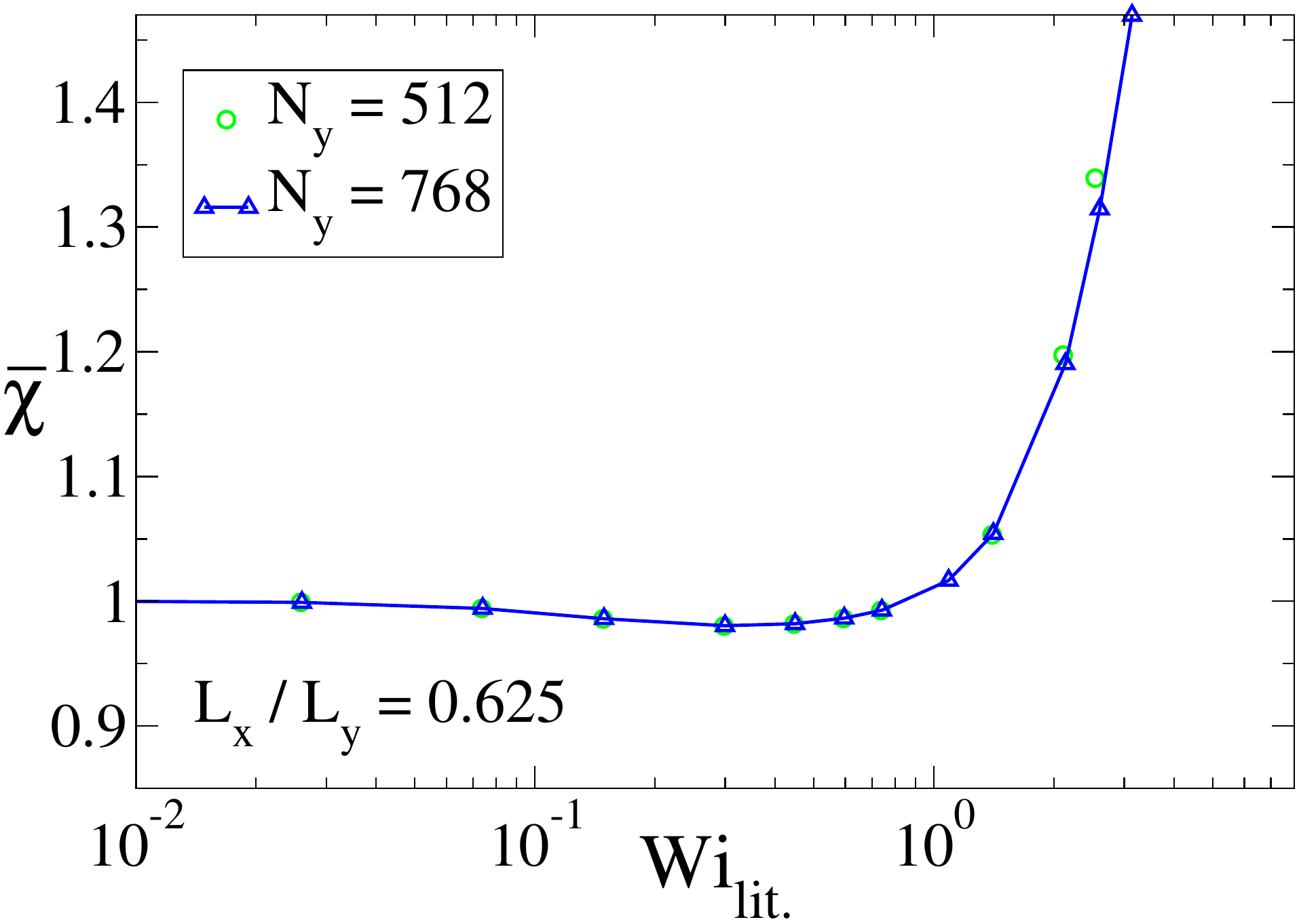}
  \caption{Cylinder drag $\bar{\chi}$ for widely spaced (left) and closely spaced (right). Data are for the Oldroyd-B model at two resolutions ($N_y = 512, 768$), with $R/L_y = 0.25$. All states are time-independent. }
  \label{fig:drag_Qc_Qd}
\end{figure*}

We now present our results for the channel geometry sketched in Fig.~\ref{fig:geometries} (lower), comprising a periodic linear array of cylinders bounded by solid walls. This has been studied widely in the existing literature \cite{Grilli2013,Liu1998a,Smith2003,Oliveira2005,Sahin2008,Vazquez-Quesada2012,Sahin2013}.  To allow a comparison between our results and some of those earlier studies we fix the cylinder radius $R/L_y = 0.25$, as in Refs.  \cite{Liu1998a,Vazquez-Quesada2012}. This leaves the horizontal distance $L_x$ between cylinder centres as the geometrical parameter to be varied numerically.  Noting that $L_x = 2 R=0.5L_y$ corresponds to touching cylinders and $L_x \to \infty$ to the limit of a single cylinder, we shall present results for two cases: closely spaced cylinders with $L_x/L_y = 0.625$, and widely spaced cylinders with $L_x/L_y = 1.5$.

In simulating this channel geometry we impose the pressure drop $\Delta P$ and measure the resulting flux $Q$. From these quantities, we can then calculate the normalised drag coefficient in Eqn.~\ref{eqn:dragnormed}. In this walled geometry, the pressure drop equates to a sum comprising {\em two} contributions: one stemming from the drag on the cylinder plus another stemming from the drag on the wall.  However, many earlier works in the literature report only the contribution from the drag on the cylinder. To allow a direct comparison with those works, we shall report a modified normalised drag coefficient $\bar{\chi}$, removing from Eqn.~\ref{eqn:drag} the contribution to the pressure drop stemming from the force on the wall. We adopt from herein the literature definition $\We_{lit.} = \tau V / R$.

Flowmaps for the case of widely spaced cylinders with $L_x/L_y=1.5$ are shown in Fig.~\ref{fig:porous_Qc_grid}. As can be seen, the flow character as quantified by the parameter $q$ shows extensional regions fore and aft of the cylinder, regions of simple shear at the channel walls, and regions of simple shear just above and below the cylinder.  As the Weissenberg number increases from left to right across the montage, the polymer conformation tensor becomes strongly deformed in the region of extensional wake aft of the cylinder. The corresponding normalised drag coefficient in Fig.~\ref{fig:drag_Qc_Qd} (left) shows similar behaviour as a function of Weissenberg number as in the biperiodic geometry, with an initial downturn then large upturn. This was reported also in the earlier studies of~\cite{Liu1998a,Vazquez-Quesada2012}.

Flowmaps for the case of closely spaced cylinders with $L_x/L_y=0.625$ are shown in Fig.~\ref{fig:porous_Qd_grid}. In this regime, the region between adjacent cylinders is effectively shielded from the main flow, as can be seen in the colourmap of $\lambdad / \lambdad^{max}$ and indeed the polymer conformation tensor remains largely undeformed in this region.  The flow character as quantified by $q$ shows pronounced regions of extensional flow along the lines projecting diagonally outwards into the fluid from the cylinder centre, in the region where the fluid just starts to squeeze into (and subsequently move out of) the vertical gap between the cylinder and the channel walls.  The corresponding drag coefficient (Fig.~\ref{fig:drag_Qc_Qd}, right) shows a less pronounced downturn as a function of $\We$ than for the case of widely spaced cylinders, as also reported in Ref.~\cite{Liu1998a}.

After an initial startup transient, all of the states reported in Figs.~\ref{fig:porous_Qc_grid},~\ref{fig:porous_Qd_grid} and~\ref{fig:drag_Qc_Qd} reached a time-independent steady state. While this is consistent with most previous 2D studies \cite{LiuThesis1997,Liu1998a,Hulsen2005}, it contradicts recent 2D works \cite{Vazquez-Quesada2012,Grilli2013} which reported time-dependent, turbulent-like states in 2D simulations of an Oldroyd-B fluid, in the same geometry as considered here. We now attempt to understand this discrepancy, first by appealing to the properties of incompressible 2D flow and then by discussing in turn the other differences between the two studies that might potentially explain this, including with regards to diffusivity, inertia, and fluctuations.

In Ref. \cite{Vazquez-Quesada2012}, the authors use the rms of the time-signal $\langle v_y(t) \rangle$ (where the average is taken over all space) as an order parameter for the transition to time-dependent states. Here we show that incompressible 2D flow with no-slip boundary conditions strictly requires $\langle v_y(t) \rangle = 0$. This means that any numerical scheme based on incompressible hydrodynamics (such as that used in this paper) cannot  hope to reproduce the fluctuations of Ref.~\cite{Vazquez-Quesada2012}.  Integrating the incompressibility condition along the length of the channel gives
\begin{align*}
  \int_0^{L_x} \nabla.\vecv{v}\ dx &= 0,\\
  \int_0^{L_x}  \left[\partial_x v_x(x,y) + \partial_y v_y(x,y)\right]dx &= 0,\\
\left[ v_x(L_x,y) - v_x(0, y)\right ]  + \int_0^{L_x} \partial_y v_y(x,y)dx &= 0.
\end{align*}
The left terms disappear due to periodic boundary conditions (alternatively because the flux cannot vary with $x$), leaving
\begin{align*}
 \partial_y \int_0^{L_x} v_y(x,y)dx &= 0\\
 \int_0^{L_x} v_y(x,y)dx &= C
\end{align*}
where $C$ is a constant independent of $y$. This integral must be zero at the boundaries (because $v_y = 0$ at the walls) so $C = 0$, meaning
\begin{align*}
 \int_0^{L_x} v_y(x,y)dx &= 0 \\
 \langle v_y \rangle = \int_0^{L_x} \int_0^{L_y} v_y(x,y)dx dy &= 0.
\end{align*}

This can easily be generalised to include solid obstacles such as a cylinder, giving the same result. Therefore for incompressible hydrodynamics, $\langle v_y \rangle$ is a quantity that should be exactly zero, and not used as an order parameter. A possible explanation for the results of Ref.~\cite{Vazquez-Quesada2012} could be that the flow in that case was slightly compressible, and the observed fluctuations (above Weissenberg numbers of order 1) were in the density.

\begin{figure}[t]
  \centering
  \includegraphics[width=\columnwidth]{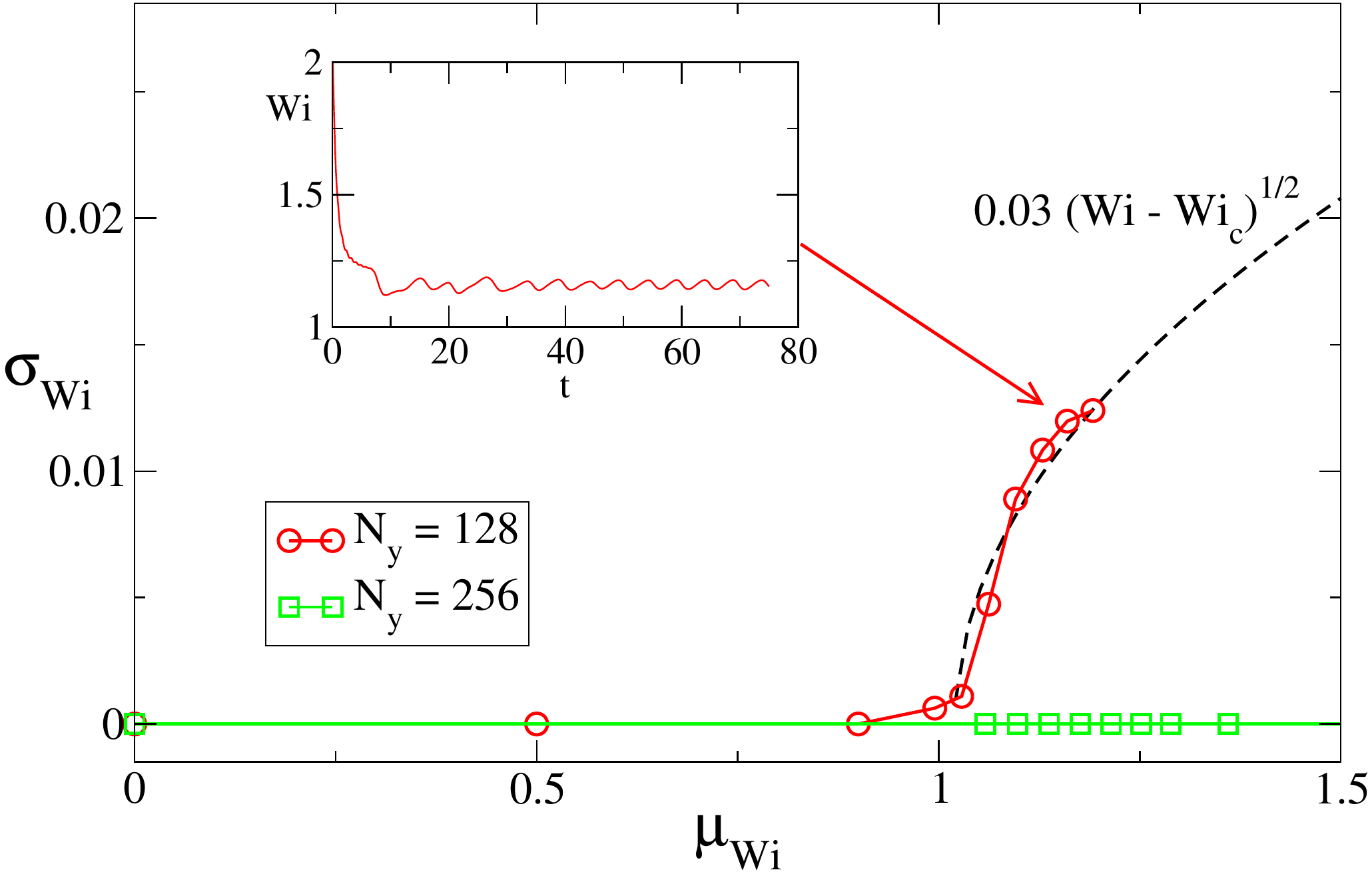}
  \caption{Plot of the standard deviation of the time series $\sigma_{\We}$ (after discarding the initial transient) against the mean $\mu_{\We}$ for two low grid resolutions, for the channel geometry with $L_x/L_y = 0.625$, $R / L_y = 0.25$. For the lowest resolution, the magnitude of fluctuations is described by the function $0.03 (\We - \We_c)^{1/2}$, where $\We_{c} \sim 1$ (black dashed line). Inset: example time-series for a run with fluctuations ($N_y = 128$). }
  \label{fig:time_dep}
\end{figure}

The above analysis only pertains to fluctuations in the $y$-component of the velocity: fluctuations in the flux (and therefore $\We$) are permitted, as are time-dependent cylinder drag or lift forces. We now demonstrate that these allowable fluctuations only arise in our simulations if insufficient numerical resolution is used. (Recall that we shall also return below to discuss several other possible sources of the discrepancy between our work and that of Ref. \cite{Vazquez-Quesada2012}.)  Focusing on the closely spaced geometry with $L_x/L_y = 0.625$, which produced strongly fluctuating states in Ref.~\cite{Vazquez-Quesada2012}, we repeat the simulations shown in \figref{fig:drag_Qc_Qd} at reduced numerical resolution. In \figref{fig:time_dep} we plot the standard deviation of the time series $\We(t)$ against the mean (after discarding the startup transient). For lowest resolution studied, we observe the onset of apparent time-dependent behaviour at $\We = \We_{c} \sim 1$.  Similar to the results of Ref.~\cite{Vazquez-Quesada2012}, the magnitude of the fluctuations approximately scales as $0.03 (\We - \We_{c})^{1/2}$, suggestive of a Hopf bifurcation. For the larger resolution, $N_y = 256$, we observe no fluctuations across the full range of $\We$ shown. However for the largest value of $\We$ shown in \figref{fig:drag_Qc_Qd} (left), even $N_y = 256$ is insufficient and similar fluctuations develop (not shown); these fluctuations disappear in our highest resolution simulations at $N_y = 512$ and $N_y = 768$.

Having shown that we do not find viscoelastic turbulence in our simulations, we return to discuss the several differences between out work and that of Refs.~\cite{Grilli2013,Vazquez-Quesada2012} that might potentially explain this.  First the model presented here includes a diffusive term. As discussed by Sureshkumar and Beris \cite{Sureshkumar1995}, this can have a stabilising effect, suppressing possible numerical instabilities.  Here we have made an effort to minimise the effect of such a term by getting as close as is numerically possible to the dual limit $\Delta x, \Delta y / \ell \to 0$, $\ell / L_x, L_y \to 0$, which ensures that the diffusive lengthscale $\ell$ is well resolved yet small compared to any features of the flow field. This approach suppresses spurious numerical instabilities whilst keeping physical ones (if present).  A second explanation could be due to the presence of small but non-zero inertia in Ref.~\cite{Vazquez-Quesada2012}. As discussed by Hoda \etal \cite{Hoda2008}, weak inertial effects can provide a mechanism by which elastic disturbances can be amplified. Finally it is conceivable that the time-dependent states observed in Ref.~\cite{Vazquez-Quesada2012} are the result of a nonlinear instability. This mechanism has been proposed to explain observed instabilities in Poiseuille flow of viscoelastic fluids, which are believed to be linearly stable \cite{Bertola2003,Pan2013}. A final possibility concerns spatial resolution. While the number of SPH particles cannot necessarily be directly compared to the number of finite-difference grid points, we note that the second largest particle resolutions in Ref. \cite{Vazquez-Quesada2012} would roughly correspond to the lowest resolution grid in \figref{fig:time_dep} for which we see fluctuations.

\section{Conclusions}

In this work, we first studied two dimensional creeping flow of the Oldroyd B, FENE-CR and FENE-P viscoelastic fluids past a biperiodic square array of cylinders. Our aim has been to understand the dramatic upturn reported experimentally in the drag coefficient as a function of increasing Weissenberg number.

By performing simulations across a wide range of values of the
porosity of the flow geometry, from `dilute' to near-touching
cylinders, we have demonstrated two qualitatively different mechanisms
that may separately underpin this thickening effect. The first
operates in the highly porous case of widely spaced cylinders, and
involves a strong stretching of the polymer chains in the extensional
wake aft of each cylinder. The second operates in the regime of more
densely packed obstacles, and involves a strong deformation of the
polymer chains as they squeeze into the vertical gap between
vertically adjacent cylinders. Two different Weissenberg numbers
separately characterise each of these regimes, and we have
demonstrated that the upturn in the drag coefficient occurs at a fixed
value of the maximum of these two numbers across the full range of
medium porosities.  We have also studied the creeping flow of an Oldroyd B fluid past a linear array of cylinders confined to a channel bounded by solid walls, where we have found that the flow remains steady for all Weissenberg numbers explored.

All the simulations in this work have assumed a purely two-dimensional flow, with translational invariance into the page in the sketches of Fig.~\ref{fig:geometries}. In experimental practice, the upturn in the drag coefficient is often accompanied by the onset of three dimensional flows (which are furthermore often time-dependent). To capture these effects, we would need to perform simulations in fully three dimensions. The two dimensional results reported here should clearly be treated with caution for Weissenberg numbers exceeding $O(1)$, where three dimensional effects may pertain.

James \cite{James2016} recently considered the ratio of shear- to extensionally-induced first normal stresses in viscoelastic flow past a biperiodic array of cylinders. They demonstrated an $O(1)$ lower bound on this, implying that shear generated first normal stresses $N_1$ cannot be neglected.  Wagner and McKinley~\cite{Wagner2016} recently examined the response of an Oldroyd-B fluid to a flow with a character-parameter $\alpha(t)$ varying sinusoidally in time, intended to mimic the variation experienced by a given fluid element as it moves through a periodic flow cell of the kind studied in this work.  For large Deborah numbers $De$ (defined as the ratio of the polymer relaxation time to the time required to pass one unit cell), they demonstrated that attempts to predict the first normal stress $N_1$ just above and below the cylinder using the \textit{local} Weissenberg number fails to reflect the previous flow history and drastically underestimates the value of $N_1$. In view of these recent works, in future numerical studies of the kind performed here it would clearly be interesting to consider more explicitly the role of normal stresses.

\section*{Acknowledgements}

EJH and SMF gratefully acknowledge financial support for this work from Schlumberger Gould Research. We also thank the referees for their valuable comments on the paper.

\bibliographystyle{prsty}
\bibliography{porous}

\end{document}